\documentclass{article}
\usepackage{graphicx}
\usepackage{amssymb}
\usepackage{amsmath}
\begin{document}

\title{On the level spacing distribution\\ in quantum graphs
\footnote{to appear in Journal of Statistical Physics}}
\author{F. Barra and P. Gaspard\\Center for Nonlinear Phenomena and Complex
Systems, \\Universit\'{e} Libre de Bruxelles, C.P. 231,\\B-1050 Brussels,
Belgium}
\date{}
\maketitle

\begin{abstract}
We derive a formula for the level spacing probability distribution in
quantum graphs. We apply it to simple examples and we discuss its 
relation with previous work and its possible application in more  general 
cases.
Moreover, we derive an exact and explicit formula for the level spacing 
distribution
of integrable quantum graphs.

\end{abstract}

\section{Introduction}

One of the major discoveries in the field of quantum chaology is the existence
of universal statistical fluctuations in the spectrum of systems that are
classically chaotic in the limit $\hbar\rightarrow0$.  These statistics are
well described by random matrix theory (RMT) in which the Hamiltonian of the
specific system under consideration is replaced by a matrix where each element
is an independent random variable except for global symmetries required by the
Hamiltonian \cite{bohigas}.  Beside the universal aspects, some statistical
properties may also depend on the particular system under consideration. The 
main
tool to study all these phenomena is the Gutzwiller trace formula that gives a
semiclassical approximation to the density of states in terms of the periodic
orbits of the corresponding classical system \cite{gutzwiller}.  The application
of
this formula has satisfactorily explained some statistical properties that agree
with RMT for chaotic systems \cite{berry1}. Nevertheless, there is no
satisfactory complete explanation yet for the universal random character of
the spectrum appearing from a specific Hamiltonian.

Recently, Kottos and Smilansky have studied very simple quantum systems
called quantum graphs that display statistical spectral fluctuations
belonging to the class of systems with a chaotic classical
limit \cite{smilansky0,smilansky1}. A remarkable aspect of the quantum graphs is that
there exists an exact trace formula that expresses the density of states in
terms of the periodic orbits of the corresponding classical dynamics in a
similar way as the Gutzwiller formula does for Hamiltonian systems. These
nontrivial features of these extremely simple systems have made them 
natural toy models of quantum chaology.

In the same perspective, several papers have been very recently devoted to these systems 
\cite{smilansky2,smilansky3,smilansky4,keating}.
On the one hand, Kottos and Smilansky have studied scattering processes in quantum graphs 
showing that these systems display all the features which characterize
quantum chaotic scattering \cite{smilansky2}.
On the other hand, the analysis of the statistical spectral fluctuations on graphs has been
considered by Schanz and Smilansky \cite{smilansky4} as well as
by Berkolaiko and Keating \cite{keating}. 
These last authors have studied for star graphs the two-point correlation function,
a  quantity which reflects the long-range spectral correlations.
In their analysis they introduce ensemble averages (for example over the lengths 
of the bonds)
in order to get a formula which is exploited by a combinatorial analysis.
 
In the present article, our aim is different in two main aspects. Firstly, we want to consider
the spacing probability distribution, which reflects short-range spectral correlations
and, secondly, we want to study the dependence of this distribution on the parameters 
of the system, in particular, on the bond lengths. Accordingly, we do not introduce external average
but we develop a method based on ergodicity.
With this purposes, we derive a general
formula for the level spacing probability distribution in quantum graphs
using a  very simple ergodic theorem.
This formula applies
more generally, too every system with levels determined by the zeros of a
quasi-periodic secular equation.  The result being exact, it contains all the
information on the particular system. To obtain the
universal behavior observed in some graphs from this result, further assumptions
and simplifications should be made.  We do not address here this difficult
problem. Instead we apply our result to very simple graphs, which nevertheless
gives interesting results (such as level repulsion) and which can guide the
approach to more difficult and interesting cases.

In Section \ref{sec.elevels}, we review some results about the quantum
mechanics on graphs. In Section \ref{sec.ps}, we derive our main
result, which is a general formula for the level spacing probability
distribution given in terms of a Poincar\'e mapping defined in a certain
surface of section $\Sigma$. In Section \ref{sec.dens}, we use the density of
states for graphs to obtain information about $\Sigma$. In Section
\ref{sec.app}, we illustrate our result with some simple graphs. In Section
\ref{sec.num}, we compare the level spacing distribution obtained numerically
for a complex graph, with the result of RMT. Then, in
Section \ref{sec.berry}, we compare our result with a related theory proposed
by Berry.  Conclusions are drawn in Section \ref{sec.conc}.

\section{Energy levels of quantum graphs}
\label{sec.elevels}
In this section, we introduce the main results known about the energy levels of
quantum graphs in order to be complete. We refer to the works of Kottos and
Smilansky for details\cite{smilansky1}.

Graphs are vertices connected by bonds.  Each bond $b=(i,j)$ connects two
vertices, $i$ and $j$. On each bond $b$, the component $\Psi_{b}$ of the total
wave function $\Psi$ is a solution of the one-dimensional Schr\"{o}dinger
equation. Here we consider the time reversible case ($i.e.$ without magnetic
field)
\[
-\frac{d^{2}}{dx^{2}}\Psi_{b}(x)=k^{2}\Psi_{b}(x),\qquad b=(i,j)\ ,
\]
where $k$ is the wavenumber.  Moreover, the wave function must satisfy boundary
conditions at the vertices of each bond ($i$ and $j$ in the previous equation),
which ensures continuity and current conservation, $i.e.$,
\[
\Psi_{b}(0)=\varphi_{i}
\]
for all the bonds $b$ which start at the vertex $i$ and
\[
\Psi_{b}(l_{b})=\varphi_{j}
\]
for all the bonds $b$ which end in the vertex $j$. The length of the bond $b$
is denoted by $l_{b}$ or $l_{(i,j)}$. The current conservation reads
\[
\sum\nolimits^{\prime}\frac{d}{dx}\Psi_{b}(x)\Big\vert_{x\rightarrow0}=\lambda
_{i}\ \varphi_{i}
\]
where $\sum^{\prime}$ denotes a summation over all the directed bonds which
have their origin at the vertex $i$. These conditions guarantee that the
resulting Schr\"{o}dinger operator is self-adjoint.
Note that, in this formulation, each bond has two
directions and we have to distinguish between the two different directions of
a bond.  This means that the dimension of the vector $\Psi=[\Psi_{1}
(x),\ldots,\Psi_{2B}(x)]^{\rm T}$ is $2B$ where $B$ is the number of bonds of
the graph.

When $\lambda_{i}\rightarrow\infty$ (Dirichlet boundary conditions) the graph
becomes a union of noninteracting bonds. These are called ``integrable 
graphs'' because the classical dynamics corresponds to particles bouncing
in the bonds leading to a phase space with the topology of a torus. 
We come back to this case in Subsection \ref{inte.sec}.
For finite $\lambda_{i}$, the
asymptotic properties of the spectrum become independent of $\lambda$ at high
wavenumbers and, indeed, there is a convergence to the Neumann limit where all the
$\lambda$'s are equal to zero. In what follows we consider this case.

As a result of the boundary conditions, we get the secular equation which can
be written in the following equivalent ways
\begin{equation}
\det[I-S(k)]=0 \label{quant1}
\end{equation}
with $S=D(k)T$ a unitary matrix of dimension $2B$ where
\begin{equation}
D_{ab}=\delta_{ab}\ \mbox{e}^{ikl_{a}}\ ,\qquad\mbox{with }\quad l_{a}=l_{b}
\label{quant1b}
\end{equation}
and
\begin{equation}
T_{ab}=-\delta_{a\hat{b}}+\frac{2}{v^{i}} \label{quant1c}
\end{equation}
if the bonds $a$ and $b$ are connected through a vertex (here called $i)$ and
zero otherwise. The notation $\hat{b}$ defines the reverted $b$ bond.

The secular equation can also be written as
\[
\det h(k)=0
\]
where $h$ is a matrix of dimension $V$ ($V$ is the number of vertices in the
graph) given by

\begin{equation}
h_{ij}(k) =  \left\{ \begin{array}{ll}
-\sum_{m\neq i}\cot kl_{(i,m)}C_{im} & \mbox{if $\ i=j$} \\ 
(\sin kl_{(i,j)})^{-1}C_{ij} & \mbox{if $\ i\neq j$} \end{array} 
\right. 
\end{equation}
$C_{ij}$ being the connectivity matrix with elements equal to one if
the vertex $i$ is connected to $j$ and zero otherwise.

It is clear from both secular equations that the eigenvalues are given by the
zeros of an almost-periodic function.

Using Eqs. (\ref{quant1}), (\ref{quant1b}), and (\ref{quant1c}), it is possible
to write the quantization condition in terms of the zeta function
\[
\zeta(k)=\prod_{p}\left[1-\mbox{e}^{-\frac{\gamma_{p}}{2}n_{p}}\ \mbox{e}
^{i(kL_{p}+\mu_{p}\pi)}\right]=0
\]
where $p$ denotes a periodic orbit, $n_{p}$ is its period, $L_{p}$ is its
length, $\gamma_{p}$ is related to the stability of the orbit and $\mu_{p}$ is
the analogue of the Maslov index. Note that $L_{p}=\sum_{i}m_{i}l_{i}$ where
the $m_{i}$ are integer numbers. If we define $x_{i}=kl_{i}$ we can see that
$\zeta(k)=\zeta(x_{1}=kl_{1},\ldots,x_{B}=kl_{B})$ with
\begin{equation}
\zeta(x_{1},\ldots,x_{B})=\prod_{p}\left[
1-\mbox{e}^{-\frac{\gamma_{p}}{2}n_{p} }\ \mbox{e}^{i\left(
\sum _{i}
m_{i}x_{i}+\mu_{p}\pi\right)}\right] \label{zeta}
\end{equation}
Note that $\zeta(x_{1},\ldots,x_{B})$ is $2\pi$-periodic in each of the
variables, so that $\zeta(k)$ is an almost-periodic function.
It can happen that the lengths of the graph are not all
incommensurate. In that case, it is convenient to define a new function
$F(x_{1},\ldots,x_{n})$ where $n$ is the number of incommensurate lengths,
which gives $\zeta(k)$ when evaluated in $x_{1}=kl_{1},\ldots, x_{n}=kl_{n}$
(here $l_{1},\ldots,l_{n}$ are the incommensurate lengths) , $i.e.$,
\[
F(x_{1}=kl_{1},\ldots,x_{n}=kl_{n})=f(k)=\zeta(k)
\]

\section{Level spacing distribution for almost-periodic functions}
\label{sec.ps}

\subsection{The level spacings as the first-return times of a Poincar\'e
mapping}

In this section, we derive the probability distribution for the spacing
between the successive zeros of an almost-periodic function $f(k)$. Let us
call $\{k_{l}\}_{l=0}^{\infty}$ the ordered solutions of $f(k)=0$.

The probability of having two successive zeros at a distance $(s,s+ds)$ is
given by
\[
P(s)ds=\lim_{K\rightarrow\infty}\frac{\#\{k_{l}\leq K:s\leq k_{l+1}-k_{l}\leq
s+ds\}}{\#\{k_{l}\leq K\}}
\]
or equivalently by
\begin{equation}
P(s)=\lim_{N\rightarrow\infty}\frac{1}{N}\sum_{l=0}^{N-1}\delta[s-(k_{l+1}%
-k_{l})] \label{prob-s}
\end{equation}

By the definition of $f(k)$, there exists a function
$F(x_{1},x_{2},...,x_{n})$ such that
\[
f(k)=F(x_{1}=kl_{1},x_{2}=kl_{2},...,x_{n}=kl_{n})
\]
where the parameters $l_{1},l_{2},...,l_{n}$ are incommensurate real numbers,
which, for the case of graphs, form the set of incommensurate lengths, and from
which all the other lengths can be obtained by linear combinations with
rational coefficients.

The function $F(x_{1},x_{2},...,x_{n})$ is periodic in each of its arguments
$x_{i}$ with a prime period $P_{i}$. Accordingly, we can consider the function
$F$ on a torus $T^{n}:0\leq x_{i}\leq P_{i}$ with $i=1,...n$.

The equation
\begin{equation}
F(x_{1},x_{2},...,x_{n})=0 \label{surface}
\end{equation}
defines a hypersurface $\Sigma$ on $T^{n}$.

The equations
\begin{equation}
\frac{dx_{i}}{dk}=l_{i}\qquad (i=1,\ldots,n) \label{flow}
\end{equation}
define a flow on this torus.  In Eq. (\ref{flow}), the wavenumber $k$ plays the role
of the time. Because of the incommensurability of the ``frequencies'' $l_{i}$ this
flow has the remarkable property of being ergodic. We will exploit this property of
dynamical systems theory to obtain the desired expression for the level
spacing probability distribution.

First, we note that each intersection of the trajectory $\{x_{i}=kl_{i}
\}_{i=1}^{n}$with the surface $\Sigma$ gives a zero
$k_{j}\in\{k_{l}\}_{l=0}^{\infty}$.  Therefore, this surface plays the
role of a Poincar\'e surface of section for the present dynamical system.

In this hypersurface of section, the flow induces a Poincar\'e map
\begin{equation}
\left\{ \begin{array}{lll}
\xi_{n+1}  &=& \phi(\xi_{n})\label{map}\\
k_{n+1}    &=& k_{n}+\tau(\xi_{n})
\end{array}\right.
\end{equation}
where $\xi_{n}$ is a point on $\Sigma$ that is mapped by the flow on
$\xi_{n+1}$ also in $\Sigma$.  These two points are the successive intersections of
the trajectory $\{x_{i}=kl_{i}\}_{i=1}^{n}$ with the surface $\Sigma$ at the times
$k_{n}$ and $k_{n+1}$ respectively. $\tau(\xi)$ is the time of first return
to the surface of section $\Sigma$.

From Eqs. (\ref{map}), we have that
\[
k_{n+1}-k_{n}=\tau[\phi^{n}(\xi_{0})]
\]
in which $\xi_{0}$ is an initial condition belonging to $\Sigma$ where the
iteration started.

Now, we can write the spacing probability distribution (\ref{prob-s}) in the
form
\begin{equation}
P(s)=\lim_{N\rightarrow\infty}\frac{1}{N}\sum_{l=0}^{N-1}\delta\left\{s-\tau
[\phi^{l}(\xi_{0})]\right\} \label{prob-s-2}
\end{equation}
The ergodicity implies that the value of the distribution
(\ref{prob-s-2}) is almost everywhere independent of the initial
condition $\xi_{0}$, so that
$\xi_{0}$ can be any point on the torus $T^{n}$ and not necessary one corresponding
to a zero. Moreover, the ergodicity implies the existence of a measure $\nu$ on
$\Sigma$ which gives the spacing probability distribution according to
\begin{equation}
P(s)=\int_{\Sigma}\nu(d\xi)\ \delta[s-\tau(\xi)] \label{P(s)}
\end{equation}

We now turn to the determination of this invariant measure $\nu$.

\subsection{The invariant measure $\nu$}

When the lengths $l_{i}$ are incommensurate, the dynamical system (\ref{flow}) is
ergodic on the torus. That is: For any measurable function $g(x_{1}
,...,x_{n})$ defined on the torus we have that
\begin{equation}
\lim_{T\rightarrow\infty}\frac{1}{T}\int_{0}^{T}g[\varphi^{t}(x_{0}
)] \ dt=\int_{T^{n}}\mu(dx) \ g(x) \label{erg-th}
\end{equation}
where $\varphi^{t}(x_{0})=lt+x_{0}$ is the flow ($\varphi^{t}$, $l$ and
$x_{0}$ are $n$-dimensional vectors) and $\mu(dx)=\frac{dx}{\left|
T^{n}\right|  }$ is the Lebesgue measure on the torus.

Let us define the function $\Delta t[\varphi^{t}(x_{0})]$ as the time of
flight of the trajectory after the last crossing of the surface of section $\Sigma$.
If the last crossing happened at $k_{n}$ then $\Delta t[\varphi^{t}(x_{0})]=t-k_{n}$.

We replace the function $g$ by
\begin{equation}
g[\varphi^{t}(\xi_{0})]=\Theta\left\{ s-\Delta t[\varphi^{t}(\xi_{0})]\right\} \ \sum
_{\{n\}}\delta(t-k_{n}) \label{fn-g}
\end{equation}
and we compute in this case the integral of the left-hand side of Eq.
(\ref{erg-th})
\[
\int_{0}^{T}g[\varphi^{t}(\xi_{0})] \
dt=\sum_{\{n\}}\int_{0}^{T}\Theta\left\{ s-\Delta t[\varphi^{t}(\xi_{0})]\right\} \
\delta(t-k_{n})\ dt
\]
We assume that there are $N$ zeros in the interval $[0,T]$ and we call them
$k_{0},...,k_{N-1}$ so that we get
\[
\int_{0}^{T}g[\varphi^{t}(\xi_{0})] \ dt=\sum_{n=0}^{N-1}\Theta[s-(k_{n+1}
-k_{n})]=\sum_{n=0}^{N-1}\Theta\left\{ s-\tau[\phi^{n}(\xi_{0})]\right\}
\]
For large values of $T$ we can consider that $T=k_{N}$ with $N$ the number of
zeros. Denoting by $\left\langle d\right\rangle$ the mean density of zeros, we have
$N=\left\langle d\right\rangle k_{N}$ and $T=\frac{N}{\left\langle
d\right\rangle }$ so that we finally get
\begin{equation}
\lim_{T\rightarrow\infty}\frac{1}{T}\int_{0}^{T}g[\varphi^{t}(\xi
_{0})] \ dt=\left\langle d\right\rangle \lim_{N\rightarrow\infty}\frac{1}{N}
\sum_{n=0}^{N-1}\Theta\left\{ s-\tau[\phi^{n}(\xi_{0})]\right\} \label{erg-eq}
\end{equation}
We recognize the cumulative function times the mean density in the right-hand
side of the expression (\ref{erg-eq}).

To compute the right-hand side of Eq. (\ref{erg-th}), we have to write $g$ as a
function of the coordinates $x$. For this purpose, we remember that
\begin{equation}
\sum_{\{n\}}\delta(t-k_{n})=\left|  f^{\prime}(t)\right|  \delta[f(t)] \ .
\label{dist-prop}%
\end{equation}
Now $f(t)=F[\varphi^{t}(\xi_{0})]$ and $f^{\prime}(t)=\nabla F[\varphi^{t}
(\xi_{0})]\cdot l$. Replacing these expressions in (\ref{dist-prop}), and
(\ref{dist-prop}) in (\ref{fn-g}), we obtain
$$g[\varphi^{t}(\xi_{0})] =\Theta\left\{ s-\Delta t[\varphi^{t}(\xi_{0})]\right\} \
\left| \nabla F[\varphi^{t}(\xi _{0})]\cdot l\right|\ 
\delta \left\{ F[\varphi^{t}(\xi_{0})]\right\}$$
from which we infer
\[
g(x)=\Theta[s-\Delta t(x)]\ \left|  \nabla F(x)\cdot l\right| \ \delta[F(x)]
\]
where $\Delta t(x)$ is the time taken by the trajectory to arrive at $x$ since
its last crossing with $\Sigma$.

Now, we compute the right-hand side of Eq. (\ref{erg-th}) which we
denote by $I$:
\begin{equation}
I=\int_{T^{n}}\mu(dx)\ g(x)=\frac{1}{\left|  T^{n}\right|  }\int_{T^{n}}dx\ g(x)
\label{I}
\end{equation}
We perform the nonlinear change of coordinates $x\rightarrow(t,\xi)$ where
$\xi$ are the $n-1$ coordinates that parametrize the surface $\Sigma$,
{\it i.e.}
\begin{equation}
x_{i}=l_{i}t+s_{i}(\xi) \label{change-coord}
\end{equation}
where the functions $s_{i}(\xi)$ are such that $F[s_{1}(\xi),...,s_{n}(\xi)]=0$.  In
the new coordinates, the equation for the surface $\Sigma$ is $t=0$ or $t=\tau
(\xi)$.  In these new coordinates, we have that
\begin{eqnarray}
& &\Delta t(x)=t\\
& &dx=J(\xi)\ d\xi\ dt
\end{eqnarray}
with the Jacobian determinant
\begin{equation}
J(\xi)=\left|
\begin{array}
[c]{ccc}
l_{1} & \cdots &  l_{n}\\
\frac{\partial s_{1}}{\partial\xi_{1}} & \cdots & \frac{\partial s_{n}
}{\partial\xi_{1}}\\
\vdots & \ddots & \vdots\\
\frac{\partial s_{1}}{\partial\xi_{n-1}} & \cdots & \frac{\partial s_{n}
}{\partial\xi_{n-1}}
\end{array}
\right|  \label{jacobiano}
\end{equation}
and $0\leq t\leq\tau(\xi)$ where $\tau(\xi)$ is the time of first return
previously introduced. In these coordinates, $I$ is given by
\[
I=\frac{1}{\left|  T^{n}\right|  }\int_{\Sigma}d\xi\ J(\xi)\int_{0}^{\tau(\xi
)}dt\ \Theta(s-t)\ \left|  \nabla F\cdot l\right| \ \delta[F(\xi,t)]
\]
The integration over $t$ can be carried out using a new variable $u$ defined through
\begin{equation}
u(t)=F(\xi,t) \label{surf}
\end{equation}
where $\xi$ is kept constant.  Differentiating with respect to $t$ gives
$\frac{du}{dt}=\nabla F\cdot l$ and we get
\[
I=\frac{1}{\left|  T^{n}\right|  }\int_{\Sigma}d\xi\ J(\xi)\int\frac
{du}{\left|  \nabla F\cdot l\right|  }\ \Theta[s-t(u)] \ \left|  \nabla F\cdot
l\right|\  \delta(u)
\]
This integral picks up the value of $t(u)$ at $u=0$. From Eq. (\ref{surf}), we
see that $u=0$ is the equation that defines $\Sigma$ and, as we noticed after
Eq. (\ref{change-coord}), there are two solutions
$t(0)=0$ or $t(0)=\tau(\xi)$ in the new coordinates. But since $t$ is the ``time of
flight'' after  the last crossing, we consider the second solution and we finally
get
\begin{equation}
I=\frac{1}{\left|  T^{n}\right|  }\int_{\Sigma}d\xi\ J(\xi)\ \Theta[s-\tau(\xi)] \ .
\label{I2}
\end{equation}

From Eqs. (\ref{erg-th}), (\ref{erg-eq}), (\ref{I}), and (\ref{I2}), we find
the cumulative function and by differentiation with respect to $s$ we obtain
the level spacing probability density
\begin{equation}
P(s)=\lim_{N\rightarrow\infty}\frac{1}{N}\sum_{n=0}^{N-1}\delta \left\{ s-\tau
[\phi^{n}(\xi_{0})]\right\} =\frac{1}{\left\langle d\right\rangle \left|
T^{n}\right|  }\int_{\Sigma}d\xi\ J(\xi)\ \delta[s-\tau(\xi)] \label{prob-mix}
\end{equation}

On the other hand, the density can also be expressed in a geometrical form.  Indeed,
starting from its definition
\[
\left\langle d\right\rangle =\lim_{T\rightarrow\infty}\frac{\#\{k_{n}<T\}}
{T}=\lim_{T\rightarrow\infty}\frac{1}{T}\int_{0}^{T}\sum_{\{l\}}\delta
(t-k_{n})\ dt
\]
and using Eq. (\ref{dist-prop}) and Eq. (\ref{erg-th}), we have
\[
\left\langle d\right\rangle =\frac{1}{\left|  T^{n}\right|  }\int_{\left|
T^{n}\right|  }dx\ \left|  \nabla F\cdot l\right| \ \delta[F(x)]
\]
Rewriting this expression in terms of the new coordinates (\ref{change-coord})
and then doing the changes of variables (\ref{surf}), we obtain
\begin{equation}
\left\langle d\right\rangle =\frac{1}{\left|  T^{n}\right|  }\int_{\Sigma}d\xi
\ J(\xi) \label{dens}
\end{equation}
Let us observe that this expression (\ref{dens}) for the density can be
obtained directly from Eq. (\ref{prob-mix}) and the normalization condition
$\int_{0}^{\infty}P(s)ds=1$.

Accordingly, we can write the spacing probability density as
\begin{equation}
P(s)=\frac{\int_{\Sigma}d\xi\ J(\xi)\ \delta[s-\tau(\xi)]}{\int_{\Sigma}d\xi
\ J(\xi)} \label{P(s)-2}
\end{equation}
which is the central result of this paper.  The expression (\ref{P(s)-2}) has a very
simple geometrical interpretation. It gives the spacing probability density as the
ratio between the flux of the probability current $l\delta[s-\tau(\xi)]$ through the
surface $\Sigma$ and the flux of the constant velocity field $l$ through the
same surface $\Sigma$.

From Eq. (\ref{P(s)-2}), we can conclude that the invariant measure $\nu$ in
(\ref{P(s)}) is given by
\[
\nu(d\xi)=\frac{d\xi\ J(\xi)}{\int_{\Sigma}d\xi\ J(\xi)} \ .
\]

\section{The density of states as a sum rule for graphs}
\label{sec.dens}
In the previous section, we derived a formula which relates the density
of zeros of an almost-periodic function to the properties of the surface of section
$\Sigma$ defined in a torus.  The dimension of the torus equals the number of
incommensurate lengths and the periodicity $P_i$ in each variable depends on the
relations between the length $l_i$ and those which are commensurable with it. 
For example, if there is a length commensurable with $l_1$, {\it i.e.},
$l_k=\frac{p}{q}l_1$ then the variable $x_1$ will have the period $P_1=2\pi q$.
In the case where the relation is of the form $l_k=nl_1$, or all the 
lengths are incommensurable, we can always
consider that $P_i=2\pi$, $\forall i$.  In what follows, we consider this to be the
case. As a consequence, we can rewrite Eq. (\ref{dens}) as

\begin{equation}
\left\langle d\right\rangle =\frac{1}{(2\pi)^{n}}\int_{\Sigma
}d\xi\ J(\xi) \label{densg}
\end{equation}
As we have already pointed out, this expression has the geometrical interpretation
of a constant flux $l$ through the surface $\Sigma$.  Because of the periodicity of
$\Sigma$ in the $n$-dimensional real space $R^{n}$, we expect that the projection of
$\Sigma$ in all of the $n$ directions covers the complete plane. (This would be false
if $\Sigma$ was a closed surface but we suppose that this is not the case.)
Therefore, if we call $\Sigma_{i}$ the projection of $\Sigma$ in the $i^{\rm th}$
direction we have
\[
\left\langle d\right\rangle =\frac{1}{(2 \pi)^{n}}\int_{\Sigma
}d\xi\ J(\xi)=\frac{1}{ (2\pi)^{n} }\sum_{i}l_i\int_{\Sigma_{i}}
ds_{i}
\]
and $\int_{\Sigma_{i}}ds_{i}=m_i(2\pi)^{n-1}$ with $m_i$ the number of 
sheets of $\Sigma$ for the projection in the $i^{\rm th}$ direction. Consequently, we
have
\begin{equation}
\left\langle d\right\rangle =\frac{1}{2\pi}\sum_i m_{i}\ l_{i}.
\label{eq.m}
\end{equation}
The number $m_i$ can be determined for each particular case 
by inspection on the quantization formula [for example Eq. (\ref{quant1})]. 

Here, our purpose is to revert the argument and use this formula to obtain the
$m_i$. This is possible because there is a general expression for the density.
For graphs, the density of states was obtained in Ref. \cite{smilansky1} using
the properties of (\ref{quant1}), (\ref{quant1b}) and (\ref{quant1c})
together with a formula for the density of states derived in the
approach of scattering quantization. The result is 
simply given by 
\begin{equation}
\left\langle d\right\rangle=\frac{L_{\rm tot}}{\pi}\ .
\label{the.density}
\end{equation}

From Eqs. (\ref{eq.m})-(\ref{the.density}), we get the desired equation 
for the $m_i$'s
\begin{equation}
\sum_{i} m_{i}\ l_{i}=2L_{\rm tot}
\label{sum.rule}
\end{equation}
Since the sum is only performed over the incommensurate lengths, forming
a basis from which all other lengths can be obtained, this equation gives all
the $m_i$ because we can write $L_{\rm tot}$ in such a basis.

We can deduce from here that when all the lengths are incommensurable there
will be two sheets in each projection.

It is possible to reduce the ``volume'' of the torus by noticing that, in 
fact, we do not need the function $F(x_1,\ldots,x_n)$ to be periodic but
the surface $\Sigma$. Since this surface is given by $F(x_1,\ldots,x_n)=0$,
the period $P_i$ with which the surface is repeated in $R^n$ is given by the
smallest of the period or anti-period of the function in the variable $x_i$
[{\it i.e.}, the values $P_i$ for which
$F(x_1,\ldots,x_i+ P_i,\ldots,x_n)=\pm F(x_1,\ldots,x_i,\ldots,x_n)$].
Note that we call it again $P_i$ but there is no risk of confusion.
Moreover, in the rest of the paper, we shall use this definition.

\section{Application to simple systems}
\label{sec.app}

\subsection{A three-bond star graph with three different bond lengths}
\label{ex.1}
\begin{figure}[ht]
\centering
\includegraphics[width=10cm]{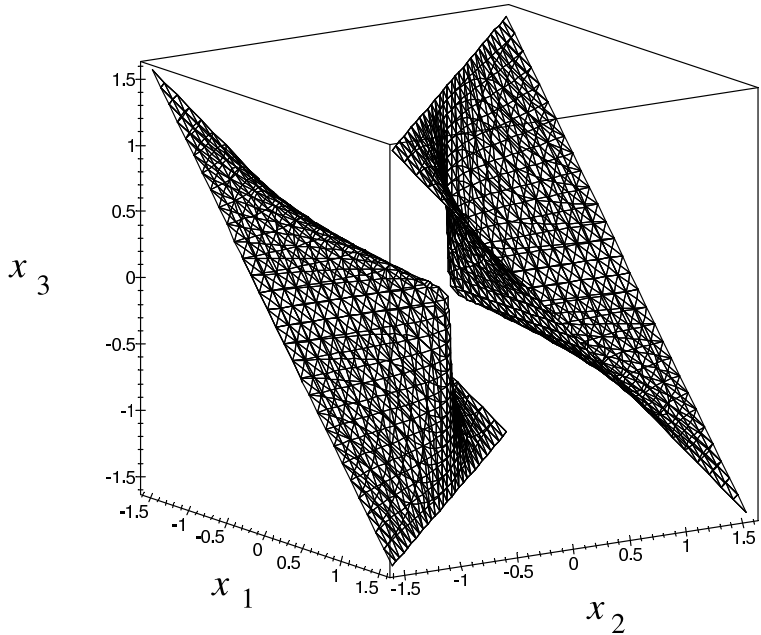}
\caption{Plot of the surface $\Sigma$ for the three-bond star graph with three
different lengths.  The plot is obtained from $G(x_1,x_2,x_3)=0$.}
\label{cone.fig}
\end{figure}

Let us consider the simple graph composed by three bonds
attached to a vertex. The spectrum of this
graph is given by the zeros of the function

\begin{equation}
f(k)=\cos kl_{1}\cos kl_{2}\sin kl_{3}+\cos kl_{1}\sin kl_{2}\cos kl_{3}+\sin
kl_{1}\cos kl_{2}\cos kl_{3} \label{ejem-3}
\end{equation}
The function (\ref{ejem-3}) is an almost-periodic function. Let us define
\[
G(x_{1},x_{2},x_{3})=\cos x_{1}\cos x_{2}\sin x_{3}+\cos x_{1}\sin x_{2}\cos
x_{3}+\sin x_{1}\cos x_{2}\cos x_{3}
\]
This function is $2\pi$-periodic in each argument but has an anti-period
$\pi$. It satisfies
\[
G(x_{1}=kl_{1},x_{2}=kl_{2},x_{3}=kl_{3})=f(k)
\]
The equation $G(x_{1},x_{2},x_{3})=0$ defines a surface $\Sigma$ with a double
cone joined by a singular point. The singular point is given by
$x_{1}=\frac{\pi}{2},x_{2}=\frac{\pi}{2},x_{3}=\frac{\pi}{2}$ 
(see Fig. \ref{cone.fig}).
For simplicity, we translate the coordinate system to that point so that we
consider the function
\[
F(x_{1},x_{2},x_{3})=G \left( x_{1}-\frac{\pi}{2},x_{2}-\frac{\pi}{2},x_{3}-\frac
{\pi}{2} \right)
\]
defined on the torus $-\frac{\pi}{2}<x_{i}\leq\frac{\pi}{2}.$

As we saw in Section \ref{sec.ps}, the shape of the surface $\Sigma$ determines the
level spacing probability distribution.  For small spacings $s$, the distribution is
given by the iterations with short ``times of flight''.  These are determined by
intersections near the singularity of $\Sigma$ because
there are arbitrarily close points in its neighborhood. In order to study the
behavior of the level spacing probability distribution for small spacings we carry
out our analysis near the singular point where the function $F$ can be approximated
by the quadratic function

\[
F(x_{1},x_{2},x_{3})=x_{1}x_{2}+x_{1}x_{3}+x_{2}x_{3}+{\cal O}(x_i^3)
\]
We diagonalize the quadratic form with a rotation of coordinates and we
finally get
\begin{equation}
F(y_{1},y_{2},y_{3})=2y_{1}^{2}-y_{2}^{2}-y_{3}^{2}+{\cal O}(y_i^3).
 \label{cono}
\end{equation}
In the $y$-coordinates, the flow is given by $\frac{dy_{i}}{dk}=e_{i}$ where
$e_{1}=(l_{1}+l_{2}+l_{3})/\sqrt{3}$, $e_{2}=(l_{2}-l_{3})/\sqrt{2}$,
$e_{3}=(l_{2}+l_{3}-2l_{1})/\sqrt{3}$. Now, we apply our theory.  We define
new coordinates through the transformation $(y_{1},y_{2}
,y_{3})\rightarrow(\eta,\xi,t)$
\begin{eqnarray}
y_{1}  &  =s_{1}(\eta,\xi)+e_{1}t\nonumber\\
y_{2}  &  =s_{2}(\eta,\xi)+e_{2}t\label{new-vars}\\
y_{3}  &  =s_{3}(\eta,\xi)+e_{3}t\nonumber
\end{eqnarray}
where the functions $s_{i}(\eta,\xi)$ are zeros of Eq. (\ref{cono}),
{\it i.e.},
$2s_{1}^{2}-s_{2}^{2}-s_{3}^{2}=0$.  A solution is
\begin{eqnarray}
s_{1}(\eta,\xi)  & =& -\sqrt{\frac{\eta^{2}+\xi^{2}}{2}}\nonumber\\
s_{2}(\eta,\xi)  & =& \eta\label{new-vars2}\\
s_{3}(\eta,\xi)  & =& \xi\nonumber
\end{eqnarray}
Eqs. (\ref{new-vars}) and (\ref{new-vars2}) define the new variables.  We need to
compute $J$ and $\tau(\eta,\xi)$. For $J$, the calculation is straightforward.
Using (\ref{jacobiano}) and (\ref{new-vars2}), we get
\begin{equation}
J=\left|  \frac{b(\eta,\xi)}{\sqrt{2(\eta^{2}+\xi^{2})}}\right|
\label{jacobian}
\end{equation}
where
\begin{equation}
b(\eta,\xi)=2e_{1}s_{1}-e_{2}s_{2}-e_{3}s_{3}=-\left[ e_{1}\sqrt{2(\eta^{2}+\xi^{2}
)}+e_{2}\xi-e_{3}\eta \right] +{\cal O}(2) \label{b}
\end{equation}
The notation ${\cal O}(2)$ means here ``to second order 
in $\eta ,\xi$ or $t$''. 
If we write Eq. (\ref{cono}) in the new coordinates we get
\begin{equation}
F(\eta,\xi,t)=\alpha^{2}t^{2}+2\ t\ b(\eta,\xi)+(2s_{1}^{2}-s_{2}^{2}-s_{3}^{2})
+{\cal O}(3)
\label{new-cone}
\end{equation}
where $b$ is defined by Eq. (\ref{b}) and
\begin{equation}
\alpha^{2}=2e_{1}^{2}-e_{2}^{2}-e_{3}^{2}=2(l_{1}l_{2}+l_{2}l_{3}+l_{1}l_{3})
\label{alpha}
\end{equation}
The third term in Eq. (\ref{new-cone}) is zero by definition. In the new
coordinates, the surface of section $\Sigma$ is given by the roots of
$F(\eta,\xi,t)=0$, {\it i.e.}, $t=0$ and $t=-\frac{2b}{\alpha^{2}}$. The function
$\tau(\eta,\xi)$ represents the ``time of flight'' of a trajectory which starts at
one point on the lower cone with coordinates $(\eta,\xi)$ and arrives to the upper
cone. That is
\begin{equation}
\tau(\eta,\xi)=-\frac{2b}{\alpha^{2}}=\frac{2\left[ \sqrt{2(\eta^{2}+\xi^{2})}
+e_{2}\xi-e_{3}\eta \right]}{2e_{1}^{2}-e_{2}^{2}-e_{3}^{2}}+{\cal O}(2)
\label{tau.1}
\end{equation}

Now we are ready to compute $P(s)$ for small $s$ using (\ref{P(s)-2}). As we
have already noticed the integral in the denominator is just the density of states
which is
\begin{equation}
\left\langle d\right\rangle =\frac{l_{1}+l_{2}+l_{3}}{\pi} \label{dens-3}
\end{equation}
for this graph.
The integral in the numerator is
\[
I=\int d\xi\ d\eta\ \frac{\left|  b(\xi,\eta)\right|  }{\sqrt{2(\xi^{2}+\eta
^{2})}}\ \delta \left(s+\frac{2b}{\alpha^{2}}\right)+{\cal O}(s^2)
\]
which is performed by changing to a variable $u(\xi)=s+\frac{2b}{\alpha^{2}}$
where $\eta$ is kept constant, by using (\ref{tau.1}), and then 
by integrating in $\eta$. The details of
this calculation are left to the reader. The result is
\[
I=\left(  \frac{\alpha}{2}\right)  ^{\frac{3}{2}}\frac{s}{\pi^{2}}
+{\cal O}(s^2)
\]
This, together with (\ref{alpha}) and (\ref{dens-3}), gives
\[
P(s)=\frac{(l_{1}l_{2}+l_{1}l_{3}+l_{2}l_{3})^{\frac{3}{2}}}{l_{1}+l_{2}
+l_{3}}\frac{s}{\pi}+{\cal O}(s^2)
\]
Usually, we express this probability density in the scaled variable $\Delta$ such
that the mean level spacing is equal to unity:
\begin{equation}
P(\Delta)=\pi\frac{(l_{1}l_{2}+l_{1}l_{3}+l_{2}l_{3})^{\frac{3}{2}}}
{(l_{1}+l_{2}+l_{3})^{3}}\Delta+{\cal O}(\Delta^2)
\label{repulsion}
\end{equation}

We observe that this simple graph already presents the Wignerian level repulsion, a
property usually associated to chaotic classical dynamics.  To our knowledge, there
are only a few systems for which this result can be derived exactly. 

\begin{figure}[ht]
\centering
\includegraphics[width=10cm]{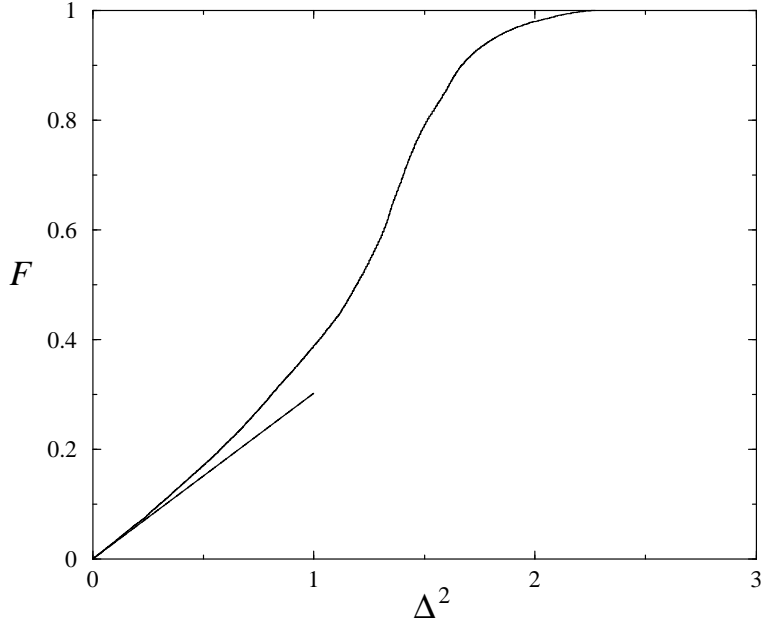}
\caption{Plot of the cumulative function $F=\int_0^{\Delta}P(\Delta^{\prime
})d\Delta^{\prime}$ of the level spacing distribution, as a function of
$\Delta^{2}$ for the three-bond star graph. The straight line is the prediction obtained by integration of
Eq. (\ref{repulsion}).  Here, $l_{1}=\pi$, $l_{2}%
=3.183459012$, and $l_{3}=3.1442336073$. }
\label{levelrep.fig}
\end{figure}

In Fig. \ref{levelrep.fig}, the cumulative function is depicted as a function
of $\Delta^2$ and compared with a numerical calculation of the spacing distribution.
The slope at the origin is half of the slope of $P(\Delta)$. The straigth line in the
figure has half of the slope given by  (\ref{repulsion}). We see that there is very
good agreement between  (\ref{repulsion}) and the numerical result.

There is an interesting point about this result. The slope of (\ref{repulsion}) 
takes values between zero and $\frac{\pi}{3^{\frac{3}{2}} }\sim 0.6$ as the lengths
$l_{1},l_{2},l_{3}$ vary.  Therefore, the slope only varies on a
relatively small interval.  This means that changing the length of the bonds (but
always keeping them irrationally related) does not change very much the slope of
the spacing probability density $P(\Delta)$. Moreover, we note that the dependence
on the lengths can be seen as a quotient between two different averages of the
lengths (a geometric average and a arithmetic one).

It is also interesting to notice in Fig. \ref{cone.fig} that the
projections of the surface $\Sigma$ onto each axis cover the corresponding
plane only once in the torus of volume $\pi^3$, which is implied by the formula
(\ref{sum.rule}) of Section \ref{sec.dens} and by Eq. (\ref{dens-3}).

\subsection{A three-bond star graph with two different bond lengths}
\label{ex.2}

\begin{figure}[ht]
\centering
\includegraphics[width=10cm]{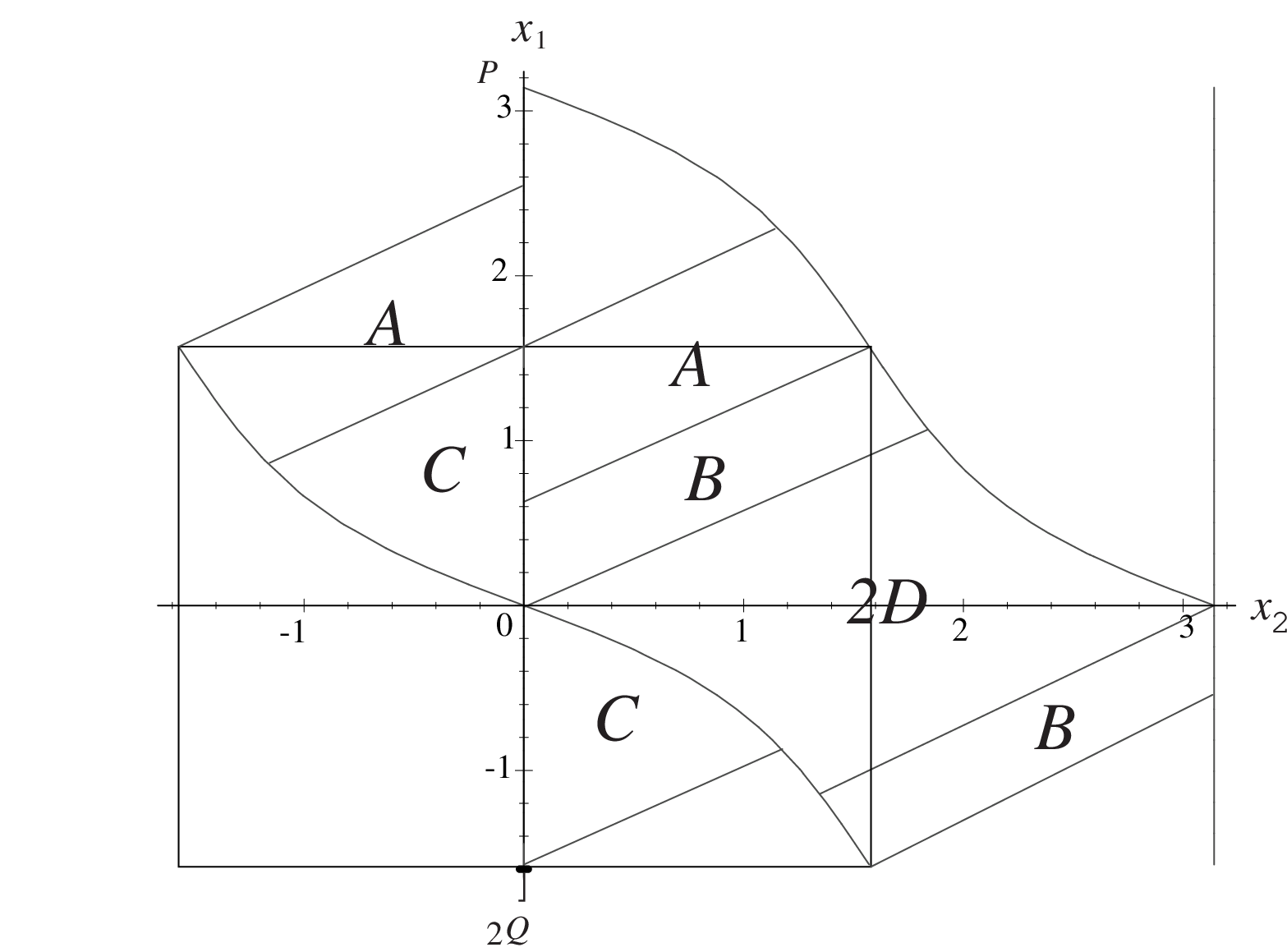}
\caption{Schematic representation of the torus for the problem
of Subsection \ref{ex.2}. The surface $\Sigma$ is composed of the $x_1$-axis and
of the curved curve. The trajectories that cross the surface $\Sigma$ belong to the
region $A,B,C$ or $D$.  Note in the figure that the regions $D$ are one next to each
other that is why we put $2D$.}
\label{reg.fig}
\end{figure}

Now, we consider the same graph as in the previous subsection but with only two
different lengths, say
$l_{1}$, $l_{2}$.  Taking $l_{1}=l_{3}$ in (\ref{ejem-3}), we get the function
which gives the zeros for this graph
\[
f(k)=\cos kl_{1} \left(  2\sin kl_{1} \cos kl_{2} +\sin kl_{2} \cos
 kl_{1} \right)
\]
We now introduce the function
\[
F(x_{1},x_{2})=\cos x_{1} \left(  2\sin x_{1} \cos x_{2} +\sin x_{2}
 \cos x_{1} \right)
\]
such that $F(x_{1}=kl_{1},x_{2}=kl_{2})=f(k).$ The function $F(x_{1},x_{2})$
is $\pi$ periodic in $x_{1}$ and $\pi$ anti-periodic in $x_{2}$ 
and can be considered in the torus
$0<x_{i}\leq\pi$ with $i=1,2$.  In Fig. \ref{reg.fig}, we draw the
lines where $F(x_{1},x_{2})=0$ in the plane $(x_{1},x_{2})$.  Changing the origin of
the coordinates to the singular point $(\frac{\pi}{2},\frac{\pi}{2})$ corresponds to
analyze the function
\[
G(x_{1},x_{2})=-\sin x_{1} \left( 2\sin x_{2} \cos x_{1} +\sin x_{1}
 \cos x_{2} \right)
\]
Note that around the singularity the function $G$ can be approximated by
the quadratic form $2x_1x_2+x_1^2$.
In this example, we explicitly obtain the density of states using $\left\langle
d\right\rangle =\frac{1}{\pi^{n}}\int_{\Sigma}J\ d\xi$. First, we identify the
surface $\Sigma$ as the union of two lines $\Sigma_{1}$ and $\Sigma_{2}$ as
indicated in Fig. \ref{reg.fig} so that
\[
\left\langle d\right\rangle =\frac{1}{\pi^{2}} \left( \int_{\Sigma_{1}}J\ d\xi
+\int_{\Sigma_{2}}J\ d\xi \right)
\]
where $\Sigma_{1}$ is given by $\sin x_{1}=0$, {\it i.e.}, $\Sigma_{1}=\{x_{1}
=0,-\frac{\pi}{2}<x_{2}<\frac{\pi}{2}\}$.  This surface is given by $s_{1}
(\xi)=0$ and $s_{2}(\xi)=\xi$ so that $J=\left|
\begin{array}
[c]{cc}
l_{1} & l_{2}\\
\frac{ds_{1}}{d\xi} & \frac{ds_{2}}{d\xi}
\end{array}
\right|  =l_{1}$. The surface $\Sigma_{2}$ is given by $2\sin x_{2} \cos
 x_{1} +\sin x_{1} \cos x_{2} =0$, {\it i.e.}, $x_{2}=-\arctan\left(\frac{1}{2}\tan
x_{1}\right)$.  Therefore, if $s_{1}(\xi)=\xi$ we find
$s_{2}(\xi)=-\arctan\left(\frac{1}{2}\tan\xi \right)$ so that we get $J=\left| 
l_{2}+\frac{2l_{1}}{1+\cos^{2}\xi}\right|  $.  Consequently, we obtain
\[
\left\langle d\right\rangle =\frac{1}{\pi^{2}} \left[ \int_{-\frac{\pi}{2}}
^{\frac{\pi}{2}}l_{1}\ d\xi+\int_{-\frac{\pi}{2}}^{\frac{\pi}{2}} \left( l_{2}
+\frac{2l_{1}}{1+\cos^{2}\xi} \right)\ d\xi\right]=\frac{2l_{1}+l_{2}}{\pi}
\]
as expected.  In Fig. \ref{reg.fig}, we notice that there are two sheets
of $\Sigma$ with projection onto $x_2$ and only one with projection onto $x_1$
as expected from the formula (\ref{sum.rule}) of Section \ref{sec.dens}.

Let us compute the level spacing probability density $P(s)$.  From the symmetry of
Fig. \ref{reg.fig}, we recognize four regions ($A,B,C,D$),
each one repeated twice.  For three of these regions ($A,B,C$), a trajectory
joins a straight line with a curved one.  For the other region ($D$), two
curved lines are joined.  In this respect, we need two expressions for the ``time of
flight'': $\tau_{1}(\xi)$ and $\tau_{2}(\xi)$.

The ``surface of arrival'' is determined by $2\sin x_{2}\cos x_{1}
+\sin x_{1}\cos x_{2}=0$, {\it i.e.}, $x_{2}=-\arctan \left( \frac{1}{2}\tan
x_{1}\right)$. Considering $x_{2}=l_{2}t+\xi$ and $x_{1}=l_{1}t$, we get
\begin{equation}
\xi=-l_{2}t-\arctan \left[ \frac{1}{2}\tan (l_{1}t) \right] \label{ejem2}
\end{equation}
Solving this equation for $t=t(\xi)$, we find that $\tau_{1}(\xi)=\left|  \min
t(\xi)\right|$.  From the periodicity of the arctangent function, the next solution
[let us call it $t_{2}(\xi)$] gives the time to cross the following surface (see
Fig. \ref{reg.fig}) so that $\tau_{2}(\xi)=t_{2}(\xi)-\tau_{1}(\xi)$.
Since the parameter $\xi$ moves in the $x_{2}$-axis, we can write
\[
\int_{\Sigma}J\ \delta[s-\tau(\xi)] \ d\xi=2l_{1}\int_{0}^{P}\delta[s-\tau_{1}
(\xi)] \ d\xi+2l_{1}\int_{Q}^{0}\delta[s-\tau_{2}(\xi)] \ d\xi
\]
where $P=\pi$ and $Q=-\frac{l_{2}\pi}{2l_{1}}$. The first integral takes the
contributions from the regions ($A,B,C$) and is easy to compute with the change of
variable $u=\tau_{1}(\xi)$. We find 
$$\int_{0}^{P}\delta[s-\tau_{1}(\xi
)]\ d\xi=\int_{0}^{\tau_{P}}\frac{du}{\left|  \frac{d\tau_{1}[\xi(u)]}{d\xi
}\right|  }\ \delta(s-u)$$ 
where $\tau_{P}=\frac{\pi}{l_{1}}-\tau_{1}(\frac
{l_{2}\pi}{l_{1}}).$ Differentiating (\ref{ejem2}) with respect to $\xi$, we
get
\[
\left|  \frac{d\tau_{1}[\xi(u)]}{d\xi}\right|  =\frac{1}{\left|  l_{2}
+\frac{2l_{1}}{1+\cos^{2}u}\right|  }
\]
so that
\[
\int_{0}^{\tau_{P}}\delta[s-\tau_{1}(\xi))] \ d\xi=\left\{
\begin{array}
[c]{c}
\left|  l_{2}+\frac{2l_{1}}{1+3\cos^{2}l_{1}s}\right|\ ,
\begin{array}
[c]{cc}
&
\end{array}
\mbox{for}\quad s<\tau_{P}\\
0\ ,
\begin{array}
[c]{cccc}
&  &  &
\end{array}
\mbox{for}\quad s>\tau_{P}
\end{array}\right.
\]
For the second integral which takes the contribution of the region $D$, we have a
similar formula but it depends on the implicit functions $\tau_{1}(\xi)$ and
$\xi(s)$ given by the equation $\tau_{2}(\xi)=s$.  If $\tau_{2}(Q)\equiv
\frac{\pi}{l_{1}}-2\tau_{1}(-\frac{l_{2}\pi}{2l_{1}})<s<\tau_{P}$ we get
\[
\int_{Q}^{0}\delta[s-\tau_{2}(\xi)]\ d\xi=\left|  \left\{ l_{2}+\frac{2l_{1}}
{1+3\cos^{2}l_{1}[s+\tau_{1}(\xi(s))]}\right\}^{-1}+\left\{
l_{2}+\frac{2l_{1}}{1+3\cos ^{2}l_{1}[\tau_{1}(\xi(s))]} \right\}^{-1}\right|  ^{-1}
\]
and zero if $s>\tau_{P}$. Hence, in the scaled variable $\Delta=\frac{L_{\rm tot}
}{\pi}s$, we have
\begin{equation}
P(\Delta)=\frac{2l_{1}}{L_{\rm tot}^{2}}\left|  l_{2}+\frac{2l_{1}}{1+3\cos
^{2}l_{1}\Delta\frac{\pi}{L_{\rm tot}}}\right|  +\frac{2l_{1}}{L_{\rm tot}^{2}}
\Gamma(\Delta) \label{pu}
\end{equation}
where $\Gamma(\Delta)=0$ if $\Delta<\frac{L_{\rm tot}}{\pi}\tau_{2}(Q)$ and
\begin{equation}
\Gamma(\Delta)=\left|  \frac{1}{l_{2}+\frac{2l_{1}}{1+3\cos^{2}l_{1}\left\{ \frac
{\pi\Delta}{L_{\rm tot}}+\tau_{1}[\xi(\Delta\frac{\pi}{L_{\rm
tot}})]\right\} }}+\frac{1}
{l_{2}+\frac{2l_{1}}{1+3\cos^{2}l_{1}\left\{\tau_{1}[\xi(\Delta\frac{\pi}{L_{\rm tot}
})]\right\} }}\right|  ^{-1} \label{pua}
\end{equation}
if $\frac{L_{\rm tot}}{\pi}\tau_{2}(Q)<s<\frac{L_{\rm tot}}{\pi}\tau_{P}$. Finally,
we note that $P(\Delta)=0$ if $\Delta>\frac{L_{\rm tot}}{\pi}\tau_{P}$.

The fact that $\Gamma(\Delta)$ has an implicit dependence on $\Delta,$ makes
difficult its actual evaluation.  Nevertheless, in the numerical example
considered below, the interval $\frac{L_{\rm tot}}{\pi}\tau_{2}(Q)<s<\frac
{L_{\rm tot}}{\pi}\tau_{P}$ where $\Gamma(\Delta)$ is different from zero is small
and we can consider a simple approximation for $\Gamma(\Delta).$

\begin{figure}[ht]
\centering
\includegraphics[width=10cm]{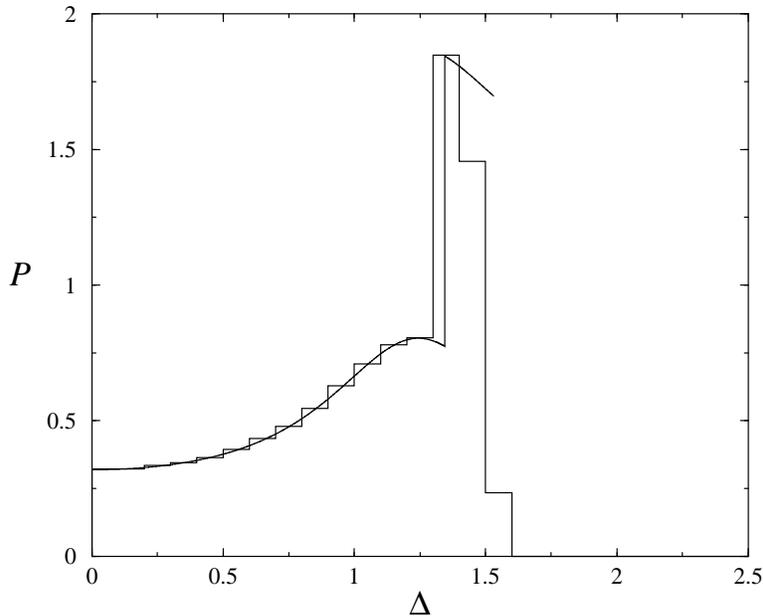}
\caption{Numerical and theoretical level spacing probability densities for
the graph of $3$ bonds with $2$ different lengths, $l_1=\pi$ and $l_2=1.53183459$.}
\label{ex2ps.fig}
\end{figure}

\begin{figure}[ht]
\centering
\includegraphics[width=10cm]{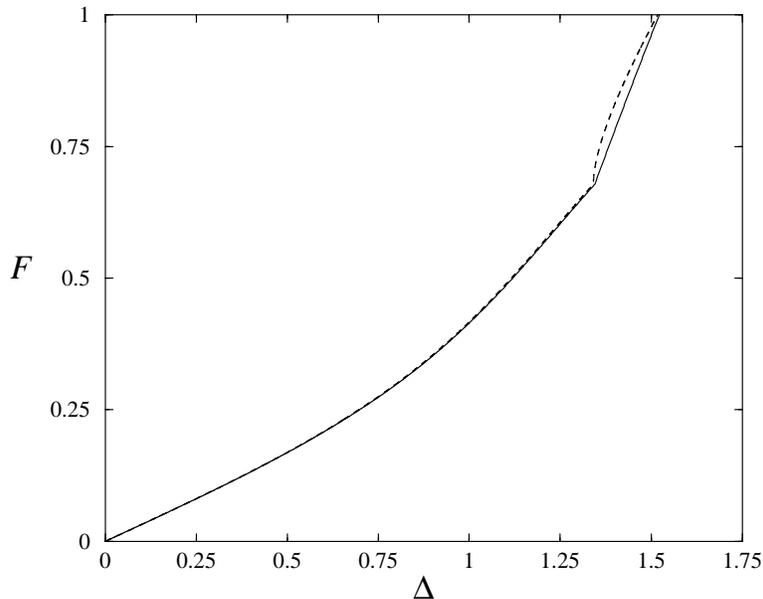}
\caption{Cumulative function for the same graph as in Fig. \ref{ex2ps.fig}.
The solid line is the theoretical calculation done 
in the text and the dashed line is the numerical result.}
\label{Fex2.fig}
\end{figure}

Consider $l_{1}=\pi$ and $l_{2}=1.53183459012.$ Thus, we get $\frac{L_{\rm tot}
}{\pi}\tau_{2}(Q)=1.345$ and $\frac{L_{\rm tot}}{\pi}\tau_{P}=1.522.$ Therefore,
$\Gamma(\Delta)$ is different from zero in the interval $1.345<\Delta<1.522$
as can be observed in Fig. \ref{ex2ps.fig}. In this interval, we can
 consider $\tau_{2}(\xi)$ as a linear function of $\xi$ (from Fig. \ref{reg.fig}
we see that the
dependence on $\xi$ is in fact smooth).  Accordingly, $\Gamma(\Delta)$ is simply
given by the constant $\frac{Q}{\tau_{2}(Q)-\tau_{p}}$. [Remember that 
$Q=-\frac{l_2\pi}{2L_{\rm tot}}$. See after Eq. (\ref{ejem2}).]  Subtituting the
numerical values, we get $\frac{2l_{1}}{L_{\rm tot}^{2}}\Gamma=1.107$
 which added to the first term of Eq. (\ref{pu}) predicts a peak of the
order of $1.8$, which agrees with the peak of the numerical result shown in
Fig. \ref{ex2ps.fig}.  A more accurate comparison can be done through the 
cumulative function $F(\Delta)=\int_0^{\Delta}d\Delta^{\prime}P(\Delta^{\prime})$. 
Fig. \ref{Fex2.fig} shows the
numerical result and the analytical result obtained by integration of
Eq. (\ref{pu}) using the approximation $\frac{2l_{1}}{L_{\rm tot}^{2}}
\Gamma=1.107$.  Here, a good agreement is observed.

\begin{figure}[ht]
\centering
\includegraphics[width=10cm]{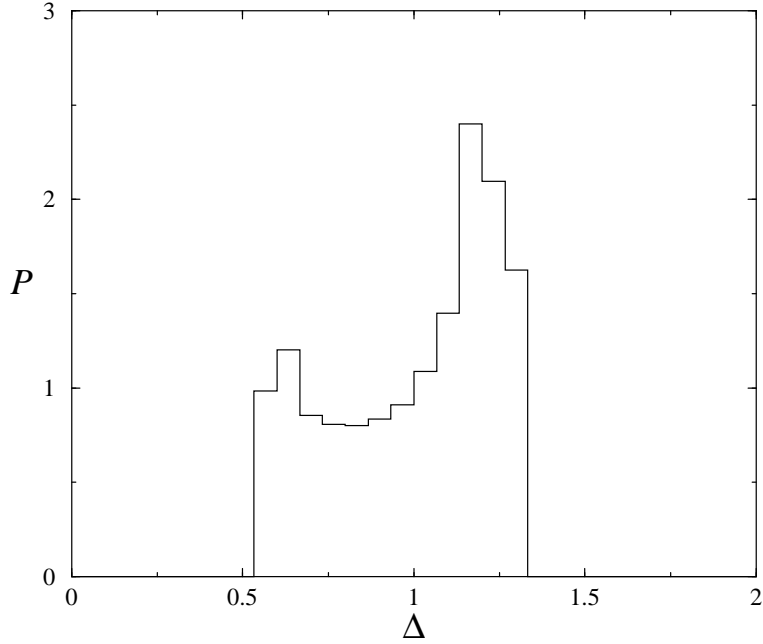}
\caption{Here, we have considered $l_3=2l_1$ in the three-bond star graph
to illustrate the existence of a critical value at which $P(\Delta)$ 
is different from zero. Here $l_1=\sqrt{2}$ and $l_2=\sqrt{3}$.}
\label{crit.fig}
\end{figure}

Let us note that if we consider this problem but with $l_3=pl_1$,
with an even integer $p$, the curve $\Sigma_1$ does not intersect $\Sigma_2$. As a
result, the level spacing probability density is zero between $\Delta=0$ and a value
$\Delta_c$. This is ilustrated in Fig. \ref{crit.fig} with $p=2$.

\subsection{Other simple graphs}
\label{other.ex}

Here, we study the level spacing in simple graphs with bonds connected to one
vertex and thus forming a loop. 

First, we consider the graph formed by two loops
attached to a single vertex. This graph has the form of an eight.
The zeros are determined by the function 
$f(k)=F(x_1=l_1k,x_2=l_2k)$ where
\begin{equation}
F(x_1,x_2)=(\cos x_2-1)\sin x_1+(\cos x_1-1)\sin x_2 \label{ex.3}
\end{equation}
This function is $2\pi$-periodic in each variable and the surface 
$\Sigma$ obtained by $F(x_1,x_2)=0$ is considered in the torus
$-\pi<x_1<\pi$ and $-\pi<x_2<\pi$. It is easy to see that this surface is
composed by $\Sigma_1=\left\{x_1=0,-\pi<x_2<\pi\right\}$, 
$\Sigma_2=\left\{x_2=0,-\pi<x_1<\pi\right\}$ and
$\Sigma_{12}=\left\{x_1+x_2=0,-\pi<x_2<\pi\right\}$.  These three sheets intersect
at the singular point $x_1=x_2=0$. The function
$F(x_1,x_2)=0$ can thus be replaced by the cubic form $x_1x_2(x_1+x_2)=0$.
The level spacing probability density can be written as
\[
P(s)=\frac{\pi}{l_1+l_2}\frac{1}{4\pi^2}\left\{\int_{\Sigma_1}J_1\ \delta[s-\tau_1(\xi)]\ d\xi+
\int_{\Sigma_2}J_1\ \delta[s-\tau_2(\xi)] \ d\xi+
\int_{\Sigma_{12}}J_{12}\ \delta[s-\tau_{12}(\xi)] \ d\xi\right\}
\]
with $J_1=l_1$, $J_2=l_2$, $J_{12}=l_1+l_2$, 
$\tau_1(\xi)=\tau_2(\xi)=\frac{\xi}{l_1+l_2}$ for $-\pi<\xi<0$, 
$\tau_1(\xi)=\tau_2(\xi)=\frac{2\pi-\xi}{l_1+l_2}$ for $0<\xi<\pi$, 
$\tau_{12}(\xi)=\frac{\xi}{l_1}$ for $0<\xi<\frac{2\pi l_{1}}{l_1+l_2}$ and
$\tau_{12}(\xi)=-\frac{\xi}{l_2}$ for $-\frac{2\pi l_2}{l_1+l_2}<\xi<0$.
Performing the integrals by using the 
variable $\Delta=\frac{l_1+l_2}{\pi}s$, we get
\begin{equation}
P(\Delta)=\left\{
\begin{array}
[c]{c}
\frac{1}{2}\ ,\qquad \mbox{if} \quad 0<\Delta<2\\ 
0\ , \qquad\mbox{otherwise} 
\end{array} \right.
\end{equation}
In this example, the spacing probability density $P(\Delta)$ is independent of the
system parameters.  We have confirmed this result with numerical calculations (data
not shown).

Another graph of a similar type is the one composed by a bond and a loop attached to
a vertex.  This graph has the form of a nine.
Here, the surface of section $\Sigma$ is given by the equation
\begin{equation}
F(x_1,x_2)=2\cos x_1 \cos x_2 - 2\cos x_1 -\sin x_1 \sin x_2 \label{ex.4}
\end{equation}
The surface can be considered in the torus $-\pi/2<x_1<\pi/2$ and 
$-\pi<x_2<\pi$ and it is given by 
\begin{eqnarray}
\Sigma_1 &=& \left\{x_2=0,-\pi/2<x_1<\pi/2\right\} \nonumber\\
\Sigma_2 &=& \left\{\tan x_1= 2(\cos x_2 -1)/\sin x_2,-\pi<x_2<\pi\right\} \nonumber
\end{eqnarray}
In this example, the calculation is similar to the one for the 
star graph with three bonds of two different lengths and we do not present it
here.  We only compute $P(\Delta)$ in the limit $\Delta \rightarrow 0$.
For the small spacings, we can consider the quadratic form around
the singularity at $(x_1=0,x_2=0)$ which is given by 
$F(x_1,x_2)\simeq x_1x_2+x_2^2$ for $x_1,x_2$ small enough. With
this approximation the calculation is similar to the one of
the previous graph.  The result is
$P(\Delta) \rightarrow \frac{l_2}{l_1+l_2}$ when $\Delta \rightarrow 0$.

\subsection{Graphs with disconnected bonds}
\label{inte.sec}
The formula (\ref{P(s)-2}) can be used to study the spacing distribution for
the ``integrable graphs'' discussed in \cite{smilansky1}. These
graphs are obtained by imposing Dirichlet boundary conditions on the vertices
and are called integrable because the classical dynamics in the graph
correspond to a particle that bounces in a bond in a periodic motion which
corresponds to a torus in phase space. In this case, the eigenvalues are
obtained by the equations
\[
\sin kl_{b}=0\ ,\qquad\forall b\ ,
\]
$i.e.$,
\[
F(x_{1},\ldots x_{n})=\prod_{i=1}^{n}\sin x_{i}=0
\]
which is the equation for the surface $\Sigma.$ This surface is composed of
all the faces of the $n$-dimensional cube which defines the torus when we
identify the corresponding boundaries, so that $\Sigma=\bigcup_{i}\Sigma_{i}$ with
$\Sigma_{i}=\{x_{i}=0,0<x_{j}<\pi\ ,\quad\forall j\neq i\}$

In this case, the level spacing probability density (\ref{P(s)-2}) is given by
\begin{equation}
P(s)=\frac{\pi}{L_{\rm tot}}\frac{1}{\pi^n}\sum_{k}\int_{\Sigma_{k}}J_{k}\ \delta[s-\tau
_{k}(s_{k})]\ ds_{k}
\label{pre.poiss}
\end{equation}
where 
\begin{eqnarray}
J_{k}&=&l_{k} \nonumber \\  \mbox{and} \qquad
\tau_{k}(s_{k})&=&\min_{j\neq k}\left\{\frac{\pi-x_{j}^{0}
}{l_{j}},\frac{\pi}{l_{k}}\right\}\ .\label{tau.poiss}
\end{eqnarray}

In the Appendix, we prove that Eq. (\ref{pre.poiss}) together with 
Eq. (\ref{tau.poiss}) are equivalent to:
\begin{equation}
P(\Delta)=\sum_{k=1}^{n}\sum_{j\neq k}^{n}\frac{l_{k}}{L_{\rm tot}}
\frac{l_{j}}{L_{\rm tot}}  \left[ \prod_{i \neq j,k}^{n}
\left( 1-\frac{l_{i}}{L_{\rm tot}}\Delta \right) \right]
\Theta \left(\frac{L_{\rm tot}}{l_{1}}-\Delta \right)+\frac{l_{1}}{L_{\rm tot}}
\left[ \prod_{i \neq 1}^{n}
\left(1-\frac{l_{i}}{l_{1}}\right)\right]\delta
\left(\Delta-\frac{L_{\rm tot}}{l_{1}}\right)
\label{qua.poiss}
\end{equation}
where $l_{1}$ is the largest length of the graph.

The distribution (\ref{qua.poiss}) is in general different
from the Poisson distribution. The Poisson distribution 
is the limit of (\ref{qua.poiss}) when $\frac{l_{1}}{L_{\rm tot}}
\rightarrow 0$. Indeed, in this limit, the delta peak vanishes,
the Heaviside function equals one and, since $\sum l_{i}=L_{\rm tot}$,
we find
$$
P(\Delta)=\lim_{n\to\infty} \prod_{i=1}^{n}
\left( 1-\frac{l_{i}}{L_{\rm tot}}\Delta \right)
= \mbox{e}^{-\Delta}
$$
Let us remark that this limit means that the number of bonds
goes to infinity but the lengths are kept constant.

\begin{figure}[ht]
\centering
\includegraphics[width=10cm]{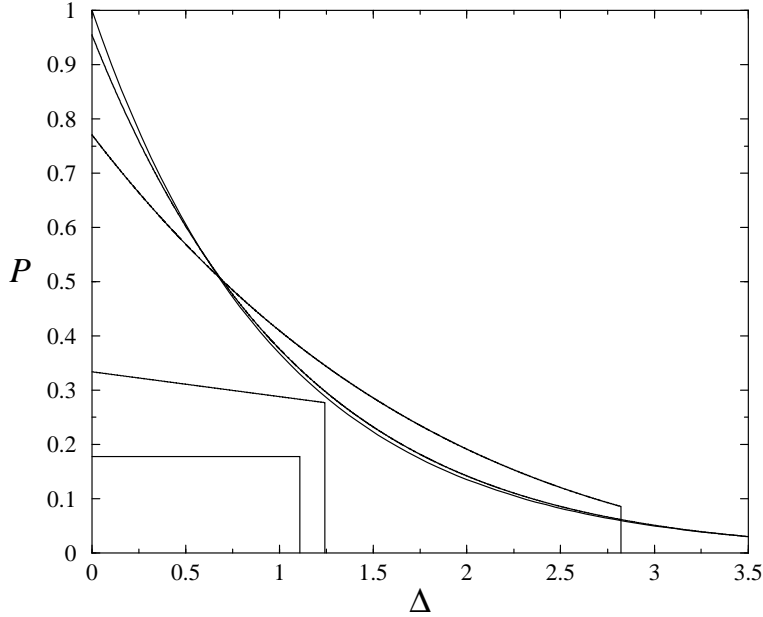}
\caption{Theoretical level spacing distribution for graphs
with disconnected bonds given by Eq. (\ref{qua.poiss}).  Here, $P$ means $P(\Delta)$. We have
omitted the delta peak from the curves. The constant distribution
is for the graph with $2$ bonds. The linear distribution is
for a graph with $3$ bonds. Then, we plot the distributions for graphs with $8$
bonds and with $30$ bonds, respectively. The last one is close to the  Poisson
distribution that is also plotted but it starts at a smaller value as predicted from
(\ref{qua.poiss}).  The lengths are given by the formula $l_{i}=\sqrt{i}$
except for $l_{1}=\sqrt{167}$, $l_{4}=\sqrt{107}$, $l_{8}=\exp(1)$,
$l_{9}=\sqrt{105}$, $l_{16}=\sqrt{119}$, and $l_{25}=\sqrt{134}$.}
\label{convergencia.fig}
\end{figure}

\begin{figure}[ht]
\centering
\includegraphics[width=10cm]{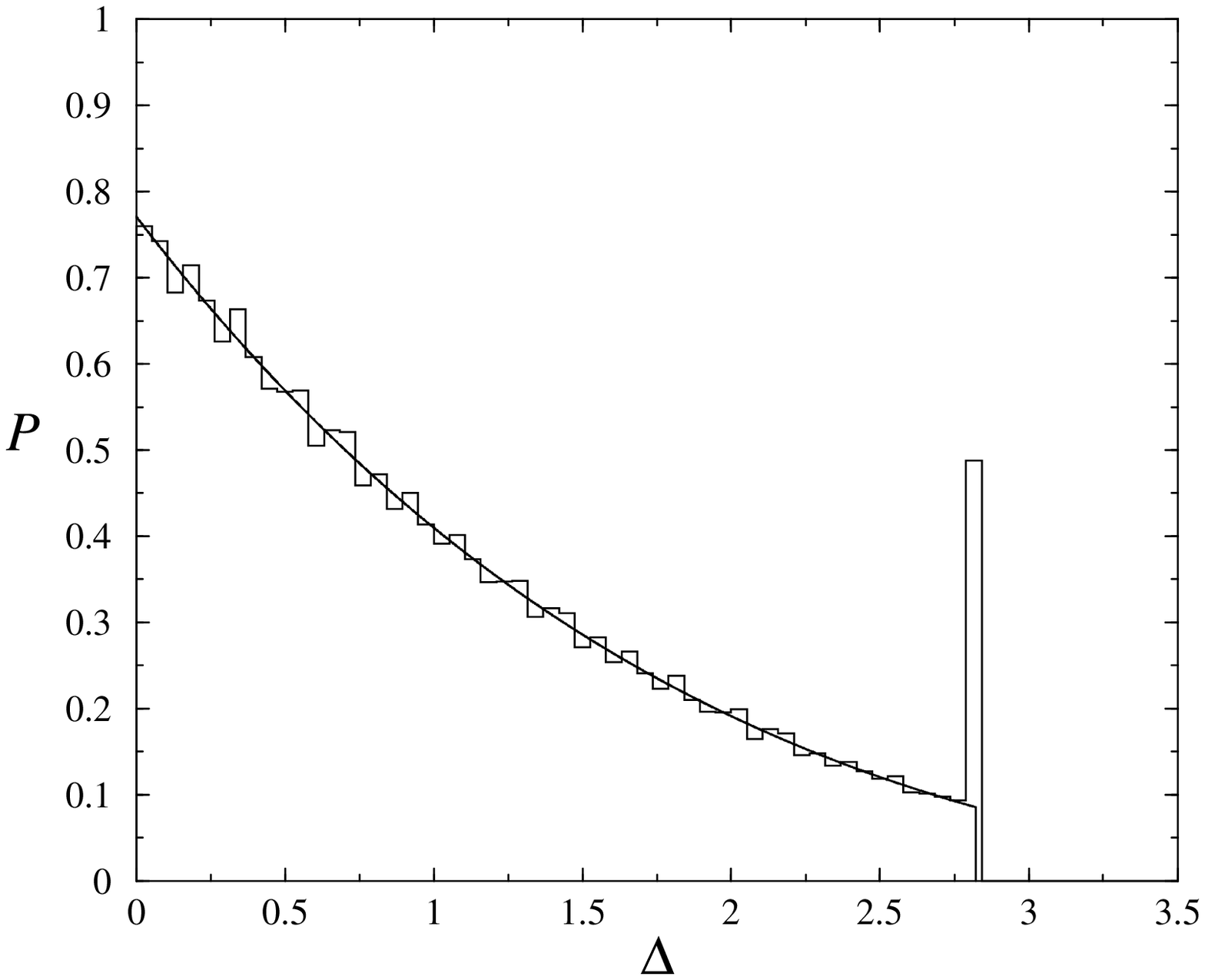}
\caption{Numerical and theoretical level spacing probability densities for a graph
with eigth disconnected bonds.  Here, $P$ means $P(\Delta)$. We have
omitted the delta peak from the theoretical curve (\ref{qua.poiss}) but we
see that its position coincides with the numerial peak.
 The histogram was built with 134050 spacings.
The lengths are $l_{1}=\sqrt{167}$, $l_{2}=\sqrt{2}$, $l_{3}=\sqrt{3}$,
$l_{4}=\sqrt{107}$, $l_{5}=\sqrt{5}$, $l_{6}=\sqrt{6}$, $l_{7}=\sqrt{7}$,
and $l_{8}=\exp(1)$.}
\label{distri8.fig}
\end{figure}

In Fig. (\ref{convergencia.fig}), we plot the distribution (\ref{qua.poiss})
for different numbers of bonds.  We observe that, for two bonds, the 
distribution is constant (except for the delta peak) and that,
for three bonds, it decays linearly. In Fig. (\ref{distri8.fig}),
we plot (\ref{qua.poiss}) and the numerical result for a graph 
of eight disconnected bonds.  We observe the very nice agreement with the formula
(\ref{qua.poiss}), as well as the convergence toward the Poisson distribution.

We want to comment on the deviations with respect to the Poisson distribution.
First, we observe a maximum spacing which is easy to understand.
The ``regular'' spectrum consists in a superposition of spectra
$\left\{\pi \frac{n}{l_{i}}\right\}$ ({\it i.e.}, equally
spaced levels). The largest spacing in this superposition 
is equal than the spacing $\frac{\pi}{l_{1}}$
where $l_{1}$ is the largest of the lengths $l$.  In the scaled variable 
of unit mean spacing, this is $\frac{L_{\rm tot}}{l_{1}}$. This maximum spacing will 
appear repeatedly over the whole $k$-axis creating the delta peak in
the distribution.

There is another interesting deviation with respect to the Poissonian distribution.
We can see from (\ref{qua.poiss}) that the probability density of finding two levels
in coincidence is
$P(0)=1-\sum_{i}\frac{l_{i}^{2}}{L_{\rm tot}^{2}}<1$.
We can compute this probability in another way:
Writing the level density $\rho(k)$ in the scaled variable $x$ of
unit mean spacing
$$
\rho(x)=\sum_{j=1}^{n}\sum_{m=0}^{\infty}
\delta \left(x-\frac{m L_{\rm tot}}{l_{j}}\right)
$$
and using the Poisson formula for the Fourier transform,
we obtain the ``power spectrum''
$$
\Pi(y)=\frac{1}{2\pi}\int_{-\infty}^{\infty}du\ \mbox{e}^{iyu}
\left<\tilde{\rho}(x)\tilde{\rho}(x+u)\right>=\sum_{j=1}^{n}
\sum_{m\neq 0}\frac{l_{j}^{2}}{L_{\rm tot}^{2}}
\delta\left(y-2\pi\frac{l_{j}m}{L_{\rm tot}}\right)
$$ 
where $\left< \right>$ is the average over $x$ and $\tilde{\rho}$
represents the fluctuations of $\rho$ around $1$ (the mean density
in this variable).
Now, the mean number of levels in the interval $[x+\Delta,x+\Delta+d\Delta]$
given that there is a level at $x$
is provided by $g(\Delta)d\Delta$ with \cite{berry-tabor}
$$
g(\Delta)=1+\int_{-\infty}^{\infty}dy\ \mbox{e}^{iy\Delta} 
\left[\Pi(y)-\frac{1}{2\pi}\right].
$$
Thus, in the case of integrable graphs, we get
$$
g(\Delta)=1-\sum_{i}\frac{l_{i}^{2}}{L_{\rm tot}^{2}}
+\sum_{j=1}^{n}\sum_{m \neq 0}
\delta\left(\Delta-\frac{m L_{\rm tot}}{l_{j}}\right).
$$
We see that $g(0)=P(0)$.
In order to compute the level spacing distribution, it is often assumed
that $P(\Delta)$ is proportional to $g(\Delta)$ and that the levels are
uncorrelated, so the probability of having two neighboring levels at a 
distance $\Delta$ is given by
$P(\Delta)=g(\Delta)\mbox{e}^{-\int_{0}^{\Delta}g(x)dx}$
\cite{berry-tabor,porter}.
We can see that these assumptions are not justified in the case of graphs
but they are approximately valid for the very small spacings and also for
the case of graphs with infinitely many bonds where the distribution is the
Poisson distribution.

\begin{figure}[ht]
\centering
\includegraphics[width=10cm]{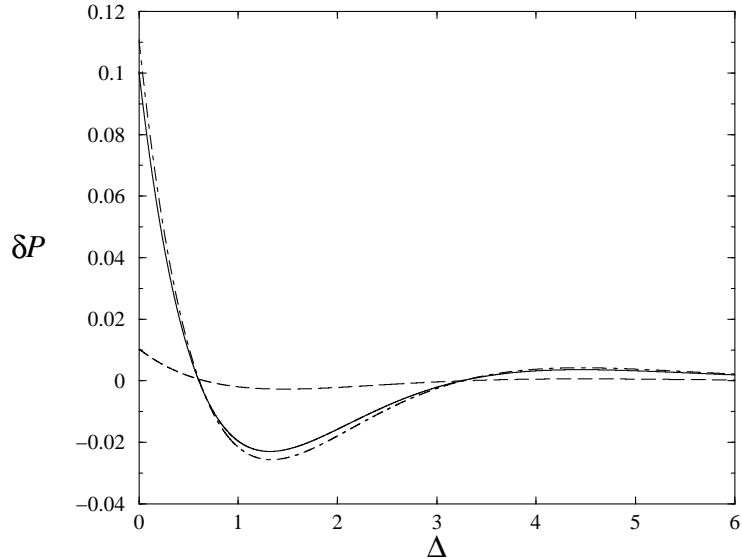}
\caption{Deviations of Eq. (\ref{qua.poiss}) with respect to
the Poisson distribution for for a graph with ten disconnected bonds.  
Here, $\delta P=\exp(-\Delta)-P(\Delta)$. 
The dot-dashed line is for the set of lengths
$l_{1}=\sqrt{3}$, $l_{2}=\sqrt{5}$, $l_{3}=\sqrt{7}$,
$l_{4}=\sqrt{11}$, $l_{5}=\sqrt{13}$, $l_{6}=\sqrt{17}$, $l_{7}=\sqrt{19}$,
$l_{8}=\sqrt{23}$, $l_{9}=\sqrt{29}$, $l_{10}=\sqrt{31}$. 
The solid line for the lengths 
$l_{1}=\sqrt{101}$, $l_{2}=\sqrt{103}$, $l_{3}=\sqrt{107}$,
$l_{4}=\sqrt{109}$, $l_{5}=\sqrt{113}$, $l_{6}=\sqrt{127}$, $l_{7}=\sqrt{131}$,
$l_{8}=\sqrt{137}$, $l_{9}=\sqrt{139}$, $l_{10}=\sqrt{149}$.
The long dashed line represents the difference between the densities evaluated with (\ref{qua.poiss})
in the two different cases.}
\label{fluct2.fig}
\end{figure}

We have explored the dependence on the lengths of the bonds in the level spacing
distribution (\ref{qua.poiss}). 
Figure \ref{fluct2.fig} shows the deviations with respect to
a Poisson distribution for two sets of lengths and the difference between them.
We see that the dependence on the lengths for a graph of 10 lengths is  very weak 
in the integrable case.
We also observe in the figure that the deviations from the Poisson distribution
is maximum for $\Delta=0$.

\section{Level spacing in complex graphs}
\label{sec.num}

\begin{figure}[ht]
\centering
\includegraphics[width=10cm]{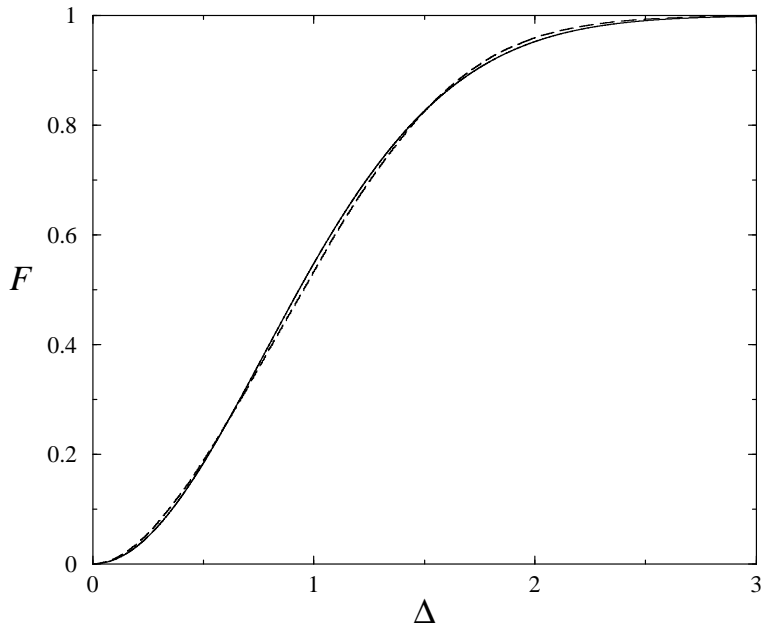}
\caption{Cumulative function of the level spacing distribution for a fully connected pentagon. 
The dashed line is the
numerical result for the pentagon with lengths $L_{i}=0.6l_{i}$ and 
$l_{1}=\sqrt{2}$, $l_{2}=\sqrt{3}$, $l_{3}=\sqrt{5}$,
$l_{4}=\sqrt{6}$, $l_{5}=\sqrt{7}$, $l_{6}=\pi$, $l_{7}=\exp(1)$,
$l_{8}=\sqrt{10}$, $l_{9}=\sqrt{11}$, $l_{10}=\sqrt{13}$.
The solid line is the RMT result.}
\label{rmt.fig}
\end{figure}

\begin{figure}[ht]
\centering
\includegraphics[width=10cm]{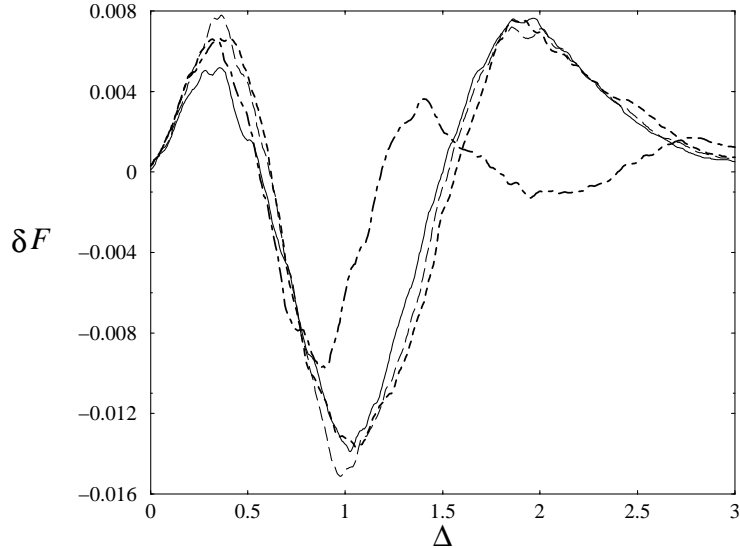}
\caption{Deviations of the cumulative function of the spacing 
distribution for different complex graphs with respect to the RMT result:
$\delta F=F-F_{RMT}$. The long dashed line represents this deviation 
for a fully conected pentagon with the same set of lengths as in figure \ref{rmt.fig}. 
The  dashed line represents this deviation for the lengths  $L_{i}=0.4l_{i}$
with the $l_{i}$ of the first set in figure \ref{fluct2.fig} 
and the solid line for the lengths $L_{i}=0.14l_{i}$ with the $l_{i}$ of the second set 
in figure \ref{fluct2.fig}. The dot-dashed line
represents this fluctuations for a tetrahedron with the lengths $L_{i}=1.05l_{i}$
$l_{1}=\sqrt{2}$, $l_{2}=\sqrt{3}$, $l_{3}=\pi$,
$l_{4}=\sqrt{6}$, $l_{5}=\sqrt{7}$, $l_{6}=\sqrt{13}$.}
\label{fluct1.fig}
\end{figure}

We have computed the level spacing distribution for a fully connected pentagon.
In figure \ref{rmt.fig} we depict the cumulative function obtained numerically
with more than 100000 levels together with the RMT prediction\cite{haake}.
 Although the agreement is 
very good some systematic deviations exist. 
In figure \ref{fluct1.fig}  we plot these deviations
for three different sets of lengths. We can see that they are very close to each other
showing that the fluctuations are independent of the graph lengths. 
The dot-dashed line in figure \ref{fluct1.fig} represents the fluctuations 
around RMT for a fully connected tetrahedron \cite{smilansky1}. 
We can conclude from these results that the fluctuations around RMT depend
on the topology of the graphs but does not depend much on their lengths. 
Moreover, we observe that the pentagon (a graph of 5 vertex, 10 bonds and valence 4) 
has bigger deviations with respect RMT than the tetrahedron 
(a graph of 4 vertex,  6 bonds and valence 3). 
This behavior is reminiscent of an observation in \cite{smilansky1} 
that for a star graph of 15 bonds the form factor deviates more from RMT that 
for a star of 5 bonds (see also \cite{keating}).
These results would suggest that the valence plays a role 
in the deviations with respect to RMT.

\section{Comparison with Berry's theory}
\label{sec.berry}
Berry has studied the level spacing distribution in classically chaotic systems with
a similar idea as the one we have developed here \cite{berry}.  He noticed that, for
a typical Hamiltonian with real eigenfunctions (which is the same situation as the one
we consider here), it is necessary to vary two parameters in order for two
levels to be degenerate.  This is the content of a theorem originally due to von
Neumann and Wigner. It also implies that, in the three-dimensional space
of the two parameters $A$ and $B$ and of the energy
$E$, the eigenvalue surface $E=E_{\pm}(A,B)$ has the
form of a double cone with its sheets joined at the ``diabolical point''
$(A^{\ast},B^{\ast},E^{\ast})$, where $A^{\ast},B^{\ast}$ are the parameters
for which the degeneracy occurs.  Following Berry, these cones are distributed in
the space $(A,B,E)$ according to a unknown probability distribution
$\rho(A,B,E)$. Berry has also considered a probability distribution which rules
the geometry of the cones [$\pi(a,b,c)$ where $a,b,c$ are the parameters in
the quadratic form which defines the cone].

The level spacing probability distribution is given by the average
(\ref{prob-s}) over energy, which can be considered in the semiclassical limit where
infinitely many levels lie near any given $E$.  As a consequence, the level spacing
is given for small spacings by the successive crossings of the conical surfaces with
the line $(E,A=A_{0},B=B_{0})$ where $A_{0}$ and $B_{0}$ are the parameters of the
actual Hamiltonian under study.  Berry argues that, since there is nothing special
about the system with the parameters $(A_{0},B_{0})$, the energy average can be
augmented by an ensemble average over a region $(A,B)$ near $(A_{0},B_{0})$. 
Whereupon, the level spacing becomes
\[
P(\Delta)=\frac{\rho(A_{0},B_{0},E)}{\langle d(E)\rangle }\int
da\, db\, dc\ \pi(a,b,c)\int dA\, dB\,\delta(\Delta-\sqrt{aA^{2}+2bAB+cB^{2}})
\]
After the change of variables $\alpha=A/\Delta$, $\beta=B/\Delta$, the previous
equation gives
\[
P(\Delta)\sim \Delta
\]
where the proportionality factor involves a geometric average.
Berry's argument shows that the level spacing density should vanish linearly
in generic systems because of the level repulsion, as expected from random matrix
theory.

The main difference between Berry's derivation and our derivation is that he
introduces by hand the ensemble average. In our derivation, the ensemble
average naturally appears from a rigorous equivalence between the energy average and
the ensemble average given by the ergodic theorem.  This ensemble average
introduced by ergodicity has the advantage of keeping all the specificities of the
system, $i.e.$, the dependence on the lengths of the graph.  We expect that these
specificities disappear for graphs which are sufficiently large, in a
way which has still to be understood for graphs with connected bonds.

\section{Conclusions and discussion}
\label{sec.conc}

In this article, we have derived a formula for the spacing probability
distribution of the energy levels of quantum graphs and, more generally, for
systems where the secular equation is given by an almost-periodic function.
Our formula is based on the ergodic properties of a continuous-time dynamical system
defined on a torus. This ergodic flow induces a
Poincar\'e map in a certain surface of section which corresponds to the locus of the
energy eigenvalues in the phase space of the flow.  The level
spacings are explicitly related to the times of first return in the surface of
section. The level spacing distribution is thus given by the distribution of the
first-return times of the ergodic flow in the Poincar\'e surface of section.  

We have applied this formula to different graphs.  In general, the slope of the
spacing density $P(\Delta)$ at $\Delta=0$ depends on the system parameters and we
have been able to calculate explicitly this dependence in several graphs. 

We have also studied in detail the ``regular'' spectrum
of integrable graphs.  One important application of our formula (\ref{P(s)-2}) 
is the following 

{\it Theorem:

If the bonds of the graph are disconnected so that the spectrum is a superposition
of $n$ independent equally spaced spectra of wavenumbers and if the bond lengths are
mutually incommensurable, the level spacing probability distribution is exactly given
by Eq. (\ref{qua.poiss}) when the distribution is expressed in the variable where the
level density is equal to one. The distribution (\ref{qua.poiss}) converges to
the Poisson distribution in the limit $n \rightarrow \infty$.}

On the other hand, for large connected graphs, the level spacing distribution
is close to the Dyson-Gaudin-Mehta spacing distribution of RMT although 
deviations are numerically observed which depend mainly on the topology
of the quantum graph. The deviations with respect to
RMT are more important for smaller graphs than for larger graphs but the effect of Wigner
repulsion is still present in very small graphs where we observe that 
the spacing density also vanishes linearly like $P(\Delta) \sim \Delta$.

The different results we have obtained can be understood on the basis of the
general properties of the surface of section $\Sigma$, which plays a particularly
important role.  First of all, we remark that the surface $\Sigma$ is defined as the
set of the zeros of $f(k) = F(kl_1,...,kl_n)=0$ in the $n$-dimensional phase space
$(x_1=kl_1,...,x_n=kl_n)$ of the ergodic flow.  Therefore, the surface
$\Sigma$ is of dimension $n-1$ in this space.  According to the von Neumann-Wigner
theorem, two zeros are generically degenerate only if two constraints are imposed
on the parameters of the systems which are here the lengths $l_i$ of the bonds. 
Consequently, the dimension of the subset of these degeneracies is $n-3$,
generically.  In the following, we refer to this subset as the singular manifold.

The aforementioned generic situation is already encountered in  graphs with three
incommensurate lengths for which the phase space of the ergodic flow is of
dimension $n=3$, the surface of section
$\Sigma$ of dimension $n-1=2$, and the degeneracy subset of dimension $n-3=0$. 
Indeed, in the example of Subsection \ref{ex.1}, the surface $\Sigma$ forms a cone
with a self-intersection at a point.  This example shows that
a spacing density vanishing linearly is generically possible as soon as there are
three incommensurate lengths.

However, for graphs with only two incommensurate bond lengths, only two
behaviors are generic: either (1) $P(0)\neq 0$ or (2) $P(\Delta)=0$ for
$0<\Delta<\Delta_c$. For such graphs, the torus is two-dimensional and the surface
giving the eigenvalue is one-dimensional, {\it i.e.}, a line on the torus. 
Generically, this line may intersect itself leading to the case (1), or it may have
no intersection leading to the case (2).  Therefore, generically, we should not
expect a spacing density which vanishes linearly like $P(\Delta) \sim \Delta$ for a
graph with only two incommensurate lengths.  This result is illustrated with the
examples of Subsections \ref{ex.2} and \ref{other.ex}.

In the two dimensional examples, we notice that
the singular manifold has the dimension $n-2$ and is a point on a two-dimensional
torus and, as a corollary, $P(\Delta)$ starts with a finite value.
This last result is supported by the fact that, for the non-generic integrable
systems, two levels can come in degeneracy by varying only one parameter. Indeed,
the singular manifold is of dimension $n-2$ for the disconnected ``integrable
graphs'' because the surface $\Sigma$ is composed by the faces of the cube and their
intersections are of dimension $n-2$, as we saw in Subsection
\ref{inte.sec}.  This discusion shows that the graphs with
two incommensurate lengths belong to a non-generic class because the singular
manifold can never be of dimension $n-3$ for $n=2$.  It is
important to notice that this statement does not contradict Berry's theory because he
considers general Hamiltonian systems where the levels are given by an equation like
${\tilde f}(k,l_1,l_2)=0$ (if we call $l_1$ and $l_2$ the two parameters that enter
in his theory) which allows the existence of a cone in the $(k,l_1,l_2)$ space,
while, for graphs with two incommensurate lengths, the secular equation has the
special form ${\tilde f}(k,l_1,l_2)=f(kl_1,kl_2)=0$ which does not allow the
existence of such a cone.

In summary, the important point which makes
the difference in the behavior of $P(\Delta)$ at small spacings $\Delta$ 
is the dimension of the singular manifold which is $n-3$ for the repulsion and a
distribution as $P(\Delta)\sim\Delta$, but $n-2$ for the clustering and a
distribution as $P(\Delta)\sim {\rm constant}$.

We notice that the degree of the polynomial that describes the surface
around the singular point seems to be not essential for this matter because
we have repulsion and clustering for cases where the surface around the
singular point is given by a quadratic form [see Subsection \ref{ex.1}
Subsection \ref{ex.2} and  Eq. (\ref{ex.4})].  We have also seen an example with
clustering [see Eq. (\ref{ex.3})] where the surface is described 
by a cubic form.

We have not commented
on the topology of the singular manifold. It can happen that the intersection
of two surfaces is transverse or tangent and this will influence
the behavior of $P(\Delta)$ near $\Delta=0$. In all the examples that we
consider, the intersection is transverse which is the generic case when there is
no restriction on the kind of surface.

Several extensions of this work are possible. 
We may wonder which are the generic properties of the surface $\Sigma$ 
and the function $\tau(\xi)$ for typical graphs and apply the formula
(\ref{P(s)-2}) to such generic situations. We notice that there are some
restrictions on $\Sigma$.  In particular, the number of sheets with projection in a
given direction depends on the number of bonds with lengths associated with this
direction, as we have seen in Section \ref{sec.dens}.  There can be other
restrictions as a consequence of the properties of the Hamiltonian (hermiticity,
etc).  In this problem, a difficulty comes from the fact that the dimension of the
torus is equal to the number of incommensurate  lengths in the graph. This makes
difficult the study of complex graphs. Another possible direction is the study of
perturbations of integrable graphs.  Such perturbations are expected to deform the
surface $\Sigma$ and we may investigate the transition from our quasi-Poissonian
distribution (\ref{qua.poiss}) to the RMT distribution. 

Since the graphs of the kind that we have discussed here have been used to model
transport in mesoscopic systems, our work can find interesting
applications in this context.

\section*{Appendix}

Here, we derive Eq. (\ref{qua.poiss}) from Eq. (\ref{pre.poiss}). The strategy
is to divide the surfaces $\Sigma_k$ into regions where the minimum, which 
appears in the definition (\ref{tau.poiss}) of $\tau_{k}(s_{k})$,
takes a given form. We call ${\cal R}_j^k$ the region where 
$\tau_{k}(s_{k})=\frac{\pi-x_j^0}{l_j}$ with $j\neq k$ and ${\cal R}^k_k$
the region where $\tau_{k}(s_{k})=\frac{\pi}{l_k}$.

First, we compute
\[
%\begin{equation}
\int_{\Sigma_{1}}ds_{1}\ \delta[s-\tau_{1}(s_{1})]=
\int_0^{\pi}dx_2^0 \cdots \int_0^{\pi}dx_n^0 \ 
\delta[s-\tau_{1}(x_2^0,\ldots,x_n^0)]
%\end{equation}
\]
with $\tau_{1}(s_{1})=\min_{j\neq 1}\left\{\frac{\pi-x_{j}^{0}
}{l_{j}},\frac{\pi}{l_{1}}\right\}$.

Consider the flow (\ref{flow}) introduced in Section \ref{sec.ps}:
\begin{eqnarray}
x_1&=&l_{1}t \nonumber \\ 
x_i&=&l_{i}t+x_i^0\ , \qquad i=2,\ldots,n 
\label{flow.1}
\end{eqnarray}

We look after any region in the surface $\Sigma_1$
where $\tau_{1}(x_2^0,\ldots,x_n^0)=\frac{\pi}{l_{1}}$.
To determine these regions, we replace $t$ by $\frac{\pi}{l_{1}}$ in (\ref{flow.1}).
This region should satisfy the following inequalities

\begin{equation}
0<x_i=l_{i}\frac{\pi}{l_1}+x_i^0 < \pi \ , \qquad i=2,\ldots,n
\label{reg.1}
\end{equation}
which express the fact that the trajectory did not cross any
boundary of the torus before arriving at $x_1=\pi$.

From (\ref{reg.1}), we get that the region ${\cal R}_1^1$ where
$\tau_{1}(s_{1})=\frac{\pi}{l_{1}}$ is given by

\begin{equation}
0<x_i^0<\pi\left(1-\frac{l_i}{l_1}\right)\ , \qquad i=2,\ldots,n
\label{R.11}
\end{equation}

We note that this result gives a border for 
all the other regions
\begin{equation}
x_i^0>\pi\left(1-\frac{l_i}{l_1}\right)\ , \qquad i=2,\ldots,n \ .
\label{border}
\end{equation}

Now, we look for the regions where 
$\tau_{1}(s_{1})=\frac{\pi-x_{j}^{0}}{l_{j}}$, $\forall j\neq 1.$
Again, from the substitution of this expression in Eq. (\ref{flow.1}), we 
obtain the following inequalities
\begin{eqnarray}
0<x_1&=&\frac{l_1}{l_j}(\pi-x_j^0)<\pi \nonumber \\ 
0<x_i&=&\frac{l_i}{l_j}(\pi-x_j^0)+x_i^0<\pi\ , \qquad i=2,\ldots,n,\quad i \neq k
\nonumber \end{eqnarray}
These inequalities imply that  
\begin{eqnarray}
\pi\left(1-\frac{l_j}{l_1}\right)<&x_j^0&<\pi \ ,\qquad j\neq 1 \nonumber \\
0<&x_m^0&<\pi-\frac{l_m}{l_j}(\pi-x_j^0)\ , \qquad \qquad m\neq \left\{1,j\right\}
\label{R.j1}
\end{eqnarray}
is the region ${\cal R}_j^1$ where 
$\tau_{1}(s_{1})=\frac{\pi-x_{j}^{0}}{l_{j}}$.

The union of ${\cal R}_1^1$ given by (\ref{R.11}) with 
${\cal R}_j^1$ ($\forall j\neq 1$) given by (\ref{R.j1})
is equal to $\Sigma_1$.

Thus, we have that
\begin{eqnarray}
\int_{\Sigma_1}ds_1\ 
\delta[s-\tau_{1}(s_1)] &=&\left[\prod_{i\neq1}^n
\int_0^{\pi\left(1-\frac{l_i}{l_1}\right)}dx_i^0 \right]
\delta \left( s-\frac{\pi}{l_1} \right)\\ &+& 
\sum_{j \neq 1}
\int_{\pi\left(1-\frac{l_j}{l_1}\right)}^{\pi}dx_j^0 \prod_{i\neq \left\{1,j\right\}}^n
\int_0^{\pi\left(1-\frac{l_i}{l_1}\right)}dx_i^0\ \delta
\left( s-\frac{\pi-x_j^0}{l_j}\right) \ . \nonumber
\end{eqnarray}
The first term represents the integration over ${\cal R}_1^1$ and the second
term the integration over ${\cal R}_j^1$. The explicit evaluation gives:

\begin{equation}
\int_{\Sigma_1}ds_1 \ 
\delta[s-\tau_{1}(s_1)]=\pi^{n-1}\left[ 
\prod_{i\neq 1}^n\left(1-\frac{l_i}{l_1}\right)\right]\delta
\left( s-\frac{\pi}{l_1} \right) +
\sum_{j \neq 1}l_j \left[\prod_{i\neq \left\{1,j\right\}}^n
(\pi-l_i s)\right]\Theta \left( \frac{\pi}{l_1}-s\right)
\label{sig.1}
\end{equation}

Now, we compute 
\begin{equation}
\int_{\Sigma_{k}}ds_{k}\ \delta[s-\tau_{k}(s_{k})]=
\int_0^{\pi}dx_1^0 \cdots \int_0^{\pi}dx_{k-1}^0 
\int_0^{\pi}dx_{k+1}^0 \cdots \int_0^{\pi}dx_{n}^0\ 
\delta[s-\tau_{k}(x_1^0,\ldots,x_{k-1}^0,x_{k+1}^0,\ldots,x_n^0)]
\end{equation}
As before, we consider the flow
\begin{eqnarray}
x_k&=&l_{k}t \nonumber \\ 
x_i&=&l_{i}t+x_i^0\ , \qquad i=1,\ldots,n ,\quad i\neq k
\label{flow.2}
\end{eqnarray}
It is easy to see that there is no region where
$\tau_{k}(s_{k})=\frac{\pi}{l_k}$. This is due to the fact that
the existence of such a region requires $\frac{l_1}{l_k}\pi+x_1^0<\pi$
and, because $l_1>l_k$, there is no positive value of $x_1^0$ where
this inequality holds.

Now, we look for the regions ${\cal R}_j^k$ where
$\tau_{k}(s_{k})=\frac{\pi-x_j^0}{l_j}$. Replacing this $\tau_{k}(s_{k})$ in
(\ref{flow.2}) we get the following conditions
\begin{eqnarray}
0<x_k&=&\frac{l_k}{l_j}(\pi-x_j^0)<\pi \\ \nonumber
0<x_i&=&\frac{l_i}{l_j}(\pi-x_j^0)+x_i^0<\pi\ , \qquad i=1,\ldots,n,\quad i \neq k
\nonumber
\end{eqnarray}
These conditions are satisfied only in the region ${\cal R}_j^k$ where 
$\tau_{k}(s_{k})=\frac{\pi-x_j^0}{l_j}$ and which is defined by
\begin{eqnarray}
\pi\left(1-\frac{l_j}{l_1}\right)<&x_j^0&<\pi \\
0<&x_m^0&<\pi-\frac{l_m}{l_j}(\pi-x_j^0)\ , \qquad m \neq \left\{j,k\right\}\ .
\end{eqnarray}
Again, the union of the regions ${\cal R}_j^k$ gives $\Sigma_k$.
Note that these regions are outside the border given by (\ref{border})
as one expected.

Thus, one has
\begin{equation}
\int_{\Sigma_{k}}ds_{k}\ \delta[s-\tau_{k}(s_{k})]=
\sum_{j\neq k}\int_{\pi\left(1-\frac{l_j}{l_1}\right)}^{\pi}
\prod_{m \neq \left\{j,k\right\}}^{n}
\int_0^{\pi-\frac{l_m}{l_j}(\pi-x_j^0)}dx_m^0\ 
\delta \left(s-\frac{\pi-x_j^0}{l_j}\right)
\end{equation}
or more explicitly
\begin{equation}
\int_{\Sigma_{k}}ds_{k}\ \delta[s-\tau_{k}(s_{k})]=
\sum_{j \neq k}^n l_j \left[\prod_{i \neq \left\{j,k\right\}}^n
(\pi-l_i s)\right] \Theta \left( \frac{\pi}{l_1}-s\right)
\label{sig.k}
\end{equation}

Finally, we substitute the results (\ref{sig.1}) and
(\ref{sig.k}) in (\ref{pre.poiss})
and we get
\begin{equation}
P(s)=\frac{l_1}{L_{\rm tot}}\left[ \prod_{i\neq 1}^n
\left(1-\frac{l_i}{l_1}\right)\right]\delta \left( s-\frac{\pi}{l_1}\right)
+\frac{1}{\pi^{n-1}L_{\rm tot}}\sum_{k=1}^n\sum_{j\neq k}^n l_jl_k
\left[\prod_{i \neq \left\{j,k\right\}}^n
(\pi-l_i s)\right] \Theta \left( \frac{\pi}{l_1}-s\right)
\end{equation}
which gives Eq. (\ref{qua.poiss}) when written in terms of the variable
$\Delta=\frac{L_{\rm tot}}{\pi}s$.

\vskip 1 cm

%%%%%%%%%%%%%%%%%%%%%%%%%%%%%%%%%%%%%%%%%%%%%%%%%%%%%%%%%%%%%%%%%%%%%%%%%%%%%%%%%
\noindent{\bf Acknowledgements.}  We dedicate this work to Professor Gr\'egoire
Nicolis on the occasion of his sixtieth birthday and we thank him for support and
encouragement in this research.  F. B. is financially supported by the
``Communaut\'e fran\c caise de Belgique" and P. G. by the National Fund for
Scientific Research (F.~N.~R.~S. Belgium).  This work is  supported, in part, by the
Interuniversity Attraction Pole program of the Belgian Federal Office of Scientific,
Technical and Cultural Affairs, by the Training and Mobility Program of the European
Commission, and by the F.~N.~R.~S. .
%%%%%%%%%%%%%%%%%%%%%%%%%%%%%%%%%%%%%%%%%%%%%%%%%%%%%%%%%%%%%%%%%%%%%%%%%%%%%%%%%

\end{document}